\documentclass[aps,prl,nofootinbib,twocolumn,superscriptaddress,preprintnumbers,balancelastpage,longbibliography]{revtex4-2}

\usepackage{placeins}
\usepackage{amsmath,amssymb,mathtools,bm}
\usepackage{graphicx, color, hepunits}
\usepackage[dvipsnames]{xcolor}
\usepackage{cancel}
\usepackage[normalem]{ulem}
\usepackage{float}
 \usepackage{filecontents}
\usepackage{multirow}
 \usepackage{hyperref}
\hypersetup{
    colorlinks=true,       
    linkcolor=blue,        
    citecolor=blue,        
    filecolor=magenta,     
    urlcolor=blue          
}
\usepackage[utf8]{inputenc}
\usepackage[english]{babel}
\usepackage{array}

\usepackage{tikz}
\usepackage{tikz-feynman}
\tikzfeynmanset{compat=1.1.0}

\newcommand{\gagg}{g_{a \gamma \gamma}}

\begin{document}

\title{
No Evidence for Superradiant Axions in LIGO-Virgo-KAGRA \\GWTC-5 Binary Black Hole Spins
}

\author{Orion Ning}
\affiliation{Leinweber Institute for Theoretical Physics, University of California, Berkeley, CA 94720, U.S.A.}
\affiliation{Theoretical Physics Group, Lawrence Berkeley National Laboratory, Berkeley, CA 94720, U.S.A.}

\author{Benjamin R. Safdi}
\affiliation{Leinweber Institute for Theoretical Physics, University of California, Berkeley, CA 94720, U.S.A.}
\affiliation{Theoretical Physics Group, Lawrence Berkeley National Laboratory, Berkeley, CA 94720, U.S.A.}

\author{Catherine Welch}
\affiliation{Leinweber Institute for Theoretical Physics, University of California, Berkeley, CA 94720, U.S.A.}
\affiliation{Theoretical Physics Group, Lawrence Berkeley National Laboratory, Berkeley, CA 94720, U.S.A.}

\date{\today}

\begin{abstract}
The quantum chromodynamics (QCD) axion and axion-like particles may form bound clouds around spinning black holes (BHs) when their Compton wavelength is comparable to the BH gravitational radius, depleting the BH spin through what is known as a \textit{superradiance} instability. Using binary BH (BBH) spin measurements obtained from the LIGO-Virgo-KAGRA GWTC-5 catalog, the most extensive public BBH catalog to date containing $N=257$ mergers with BH masses spanning roughly $5$--$135$ $M_\odot$, we perform a hierarchical Bayesian analysis in the context of a BH spin population model to constrain ultralight axions. The presence of axions at a given mass would imprint a unique signature in the observed mass-spin relation relative to the formation distribution. We find no evidence for axions across more than two decades in mass, excluding axion masses $1.7 \times 10^{-14} \, {\rm eV} \lesssim m_a \lesssim 3.3 \times 10^{-12} \, {\rm eV}$ at 95\% confidence. 
Because prior superradiance bounds in this range derive from X-ray spin measurements with substantial modeling systematics, this result represents one of the strongest robust lower bounds on the QCD axion mass.
\end{abstract}
\maketitle

\begin{figure}[tb]
\centering
\includegraphics[width=1.0\linewidth]{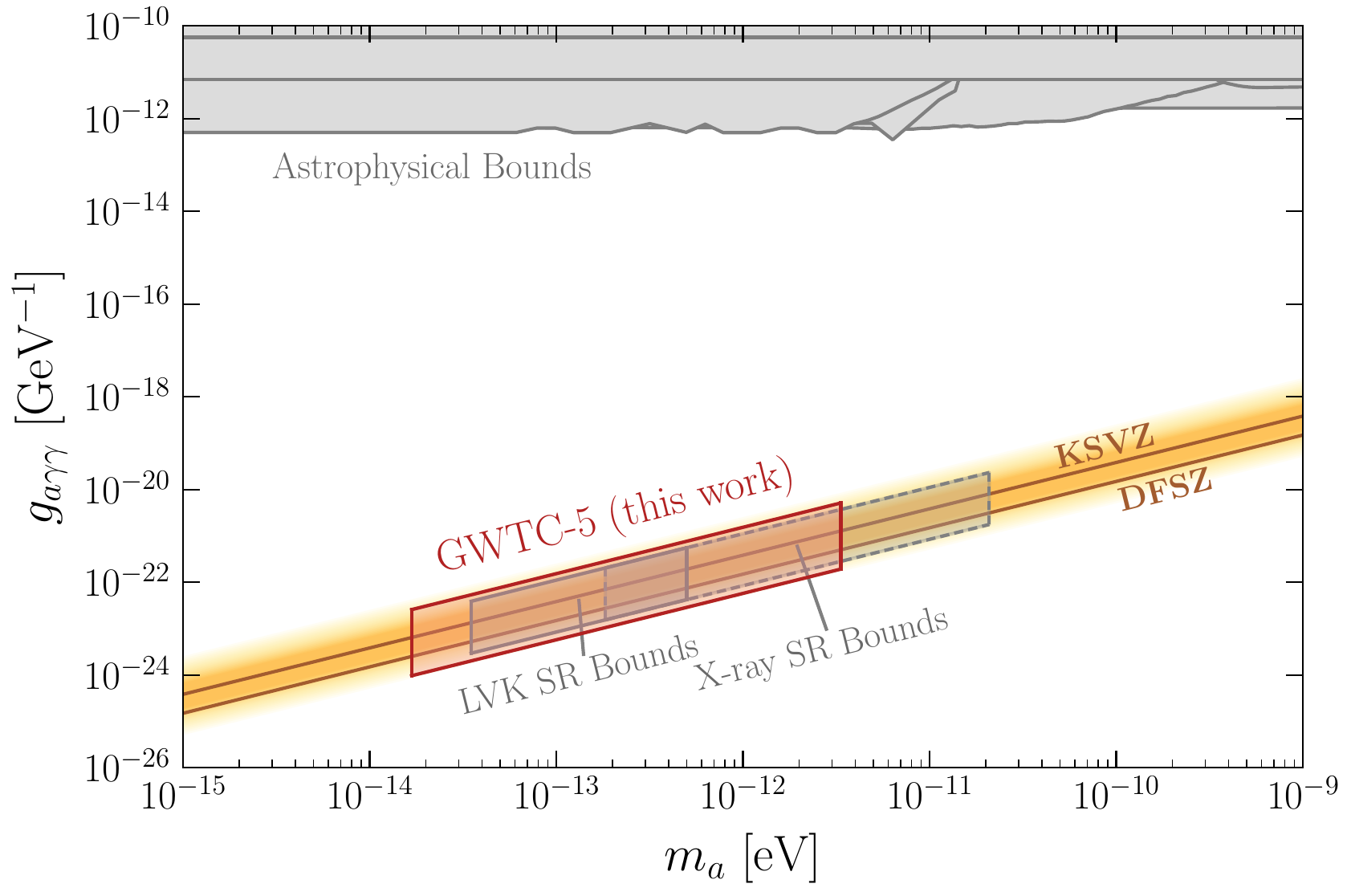}
\caption{Constraints from this work (95\% confidence exclusion region) on the QCD axion mass $m_a$ from our search for axion superradiance in the GWTC-5 BBH catalog. We illustrate our results against $\gagg$ for presentation purposes, and, since we focus on the QCD axion, truncate our constraint just above and below the QCD axion band, though our analysis does not involve $g_{a\gamma\gamma}$ directly and also applies to more general axion-like particles. 
We compare our results to existing superradiance constraints from X-ray and LVK spin measurements (gray bands), with the dashed X-ray portion indicating possible systematics from the high spins inferred for X-ray-based constraints. In the context of $g_{a\gamma\gamma}$ we also show other constraints for similar $m_a$ in gray, which are all astrophysical.}
\label{fig:gagg}
\end{figure}

\noindent
{\bf Introduction.---}\label{sec:intro}The quantum chromodynamics (QCD) axion is a well-motivated dark matter (DM) candidate that solves the Strong {\it CP} problem~\cite{Peccei:1977hh, Peccei:1977ur, Weinberg:1977ma, Wilczek:1977pj, Preskill:1982cy, Abbott:1982af, Dine:1982ah} and that emerges generically in the context of string theory compactifications~\cite{Witten:1984dg, Choi:1985je, Barr:1985hk, Svrcek:2006yi, Arvanitaki:2009fg, Demirtas:2018akl, Halverson:2019cmy, Mehta:2021pwf, Gendler:2023kjt,Benabou:2025kgx,Fallon:2025lvn,Agrawal:2025rbr,Benabou:2026jtv}.  QCD axion DM laboratory experiments around the world are rapidly scaling up~\cite{Adams:2022pbo} to search for evidence of axions in the currently-allowable axion parameter space with mass below $m_a \lesssim 20$ meV~\cite{Carenza:2019pxu,Buschmann:2021juv,Caputo:2024oqc}, with the upper bound set by neutron star (NS), supernova, and stellar cooling.  Terrestrial experiments struggle to detect QCD axions with masses below roughly $0.5$ neV~\cite{Budker:2013hfa,Kahn:2016aff,DMRadio:2022jfv} due to the axion's ultra-feeble interaction strength in this mass range.  On the other hand, axion-induced black hole (BH) superradiance (SR) is a well-established (see, {\it e.g.},~\cite{Arvanitaki:2009fg,Arvanitaki:2010sy}) and promising physical mechanism that could allow for the detection or exclusion of QCD axions at the lowest portion of the theoretically viable axion mass range, potentially up to $m_a \sim 10^{-11}$ eV or higher.  The QCD axion mass is inversely proportional to its decay constant $f_a$, $m_a \approx 0.57 \, {\rm neV} ( 10^{16} \, {\rm GeV} / f_a )$~\cite{diCortona:2015ldu}; physically we expect $f_a \lesssim m_{\rm pl}$, with $m_{\rm pl}$ the reduced Planck mass, yielding $m_a \gtrsim 3 \times 10^{-12}$ eV, though recent work from the weak gravity conjecture and the gravitational path integral suggests a slightly sharper lower bound of $m_a \gtrsim 10^{-11}$ eV~\cite{Reece:2025thc,Benabou:2025kgx,DiUbaldo:2026rly,Maldacena:2026jqd}.  In this work we produce stringent constraints on the QCD axion mass, excluding $1.7 \times 10^{-14} \, {\rm eV} \lesssim m_a \lesssim 3.3 \times 10^{-12}\, {\rm eV}$ (see Fig.~\ref{fig:gagg}) at 95\% confidence from an analysis of binary BH (BBH) spin measurements from the LIGO-Virgo-KAGRA (LVK) GWTC-5 catalog~\cite{LIGOScientific:2026wfs}. Note that our results also apply to axion-like particles and ultra-light scalars beyond the QCD axion~\cite{Arvanitaki:2009fg} as discussed more below.

Axion-induced SR proceeds as follows: when the Compton wavelength of an ultralight massive boson is comparable to the gravitational radius of a spinning Kerr BH of mass $M$, a classical wave amplification process, \textit{superradiance}, forms a hydrogen-like bound cloud around the BH~\cite{Damour1976,Ternov:1978gq,Detweiler:1980uk,Furuhashi:2004jk,Dolan:2007mj}. The cloud extracts angular momentum from the BH, depleting its spin to a critical value $\chi_c(m_a, M)$ after which SR shuts off. The result is a forbidden region in the BH mass-spin plane (the `Regge plane') above the critical-spin curve where BHs cannot persist on observationally-relevant timescales: any BH born above the curve should be rapidly spun down. This idea has been extensively  leveraged in past works, {\it e.g.}~\cite{Arvanitaki:2010sy,Arvanitaki:2014wva,Arvanitaki:2016qwi,Mehta:2020kwu,Baryakhtar:2020gao,Unal:2020jiy,Ng:2020ruv,Hoof:2024quk,Witte:2024drg,Caputo:2025oap,Aswathi:2025nxa}, to search for axions.  Our work is the first to systematically apply SR to the full LVK GWTC-5 catalog~\cite{LIGOScientific:2026wfs}, containing all events from the O1, O2, O3, O4a, and O4b observing runs.

Prior SR-based constraints on axions have been derived from
X-ray measurements of stellar-mass~\cite{Arvanitaki:2014wva,Cardoso:2018tly}
and supermassive~\cite{Stott:2018opm,Cardoso:2018tly,Mehta:2020kwu,Unal:2020jiy}
BH spins, with the stellar-mass analyses reporting exclusions in mass ranges overlapping with the analysis
presented here; for example, Ref.~\cite{Arvanitaki:2014wva} claimed to exclude axions in the mass range $6 \times 10^{-13} \, {\rm eV} < m_a < 2 \times 10^{-11} \, {\rm eV}$ using stellar mass BH spin measurements. The reliability of those constraints, however, is
fundamentally limited by the reliability of the underlying X-ray spin
measurements. Both the continuum-fitting and Fe~K$\alpha$
reflection-spectroscopy methodologies require detailed modeling of the
inner accretion disk,
and the associated
systematic uncertainties on the inferred BH spin $\chi$ are typically
comparable to or larger than the reported statistical
uncertainties~\cite{Reynolds:2020jwt}. Notably, several of the highest-spin
BHs driving the X-ray-based SR exclusions are reported with posteriors that
drive against the Kerr extremal limit at $\chi = 1$, where the pile-up may
indicate systematic mismodeling rather than genuine near-extremal rotation. For example, Ref.~\cite{Arvanitaki:2014wva} (see also~\cite{Witte:2024drg}) found that the strongest constraints on axions at high masses arise from the X-ray spin measurements of Cyg X-1 (assuming mass $M = 14.8 \pm 1.0$ $M_\odot$ and spin $\chi =0.92_{-0.18}^{+0.06}$) and  GRS 1915+105 (assuming $M = 10.1 \pm 0.6$, $\chi > 0.95$).  However, Refs.~\cite{Zdziarski:2023zuh,Zdziarski:2024zfg} claim that the Cyg X-1 spin measurement is highly model dependent and could be as low as $\chi < 0.2$ (effectively no spin at all), depending on the disk model. Similarly, Ref.~\cite{Mills:2021dxs} claims that the best-fit spin of GRS 1915+105 can vary from 0.4 to 0.99 depending on the assumed system parameters in the continuum fitting method for extracting the spin. Further doubt is raised on these types of X-ray-based spin measurements from simulation work in Ref.~\cite{Nagele:2026ppv}.

Independent indication of systematic bias in the X-ray spin measurements comes from the BBH population
inferred by LVK, whose effective spins $\chi_{\rm eff}$ are clustered near
zero in stark contrast to the high spins reported from X-ray measurements~\cite{Belczynski:2017gds,Fishbach:2021xqi}.
If the spins of the handful of BHs considered in~\cite{Arvanitaki:2014wva,Witte:2024drg} are indeed near zero, then the corresponding constraints would be significantly weakened, though it is unclear whether this is the case. On the other hand, the gravitational wave (GW)-based SR analysis we present here probes a distinctly different set of systematic uncertainties; for example, LVK parameter estimation infers BH spins from inspiral-merger-ringdown waveforms alone, with no reliance on accretion-disk modeling.

In this work we compare the ensemble of LVK-measured BH masses and spins to a BH population model that includes a natal spin distribution, age distribution, and mass distribution, in addition to the axion signal that is controlled by $m_a$. For a given $m_a$, the present-day BH distribution is modified by axion-induced SR, which suppresses the high-spin and appropriate-mass BHs and causes spin pileup near the curve $\chi_c(m_a,M)$ as a function of $M$. 
Early projections for LIGO were discussed in Ref.~\cite{Arvanitaki:2016qwi}, though more recently Ref.~\cite{Ng:2020ruv} performed this analysis on the GWTC-2 catalog, excluding axions in the mass range between roughly $1.3 \times 10^{-13}$ eV and $2.7 \times 10^{-13}$ eV; going to the GWTC-4 catalog increases the statistics to $N =153$, while the GWTC-5 catalog, which includes the O4b catalog and also a reanalysis of some older events, increases the statistics to $N=257$ events. The larger data set increases the excluded mass range by over an order of magnitude compared with the GWTC-2 results.  Our work is the first to systematically analyze all events in the GWTC-5 catalog (see also Refs.~\cite{Caputo:2025oap, Aswathi:2025nxa, LIGOScientific:2025brd}, which examined individual events from recent GWTC catalogs, and Refs.~\cite{Arvanitaki:2014wva,Arvanitaki:2016qwi,Brito:2017wnc,Tsukada:2018mbp,Palomba:2019vxe,Tsukada:2020lgt,Zhu:2020tht,Yuan:2021ebu,LIGOScientific:2021rnv,Sprague:2024lgq,Ellis:2026gwt,Yang:2017lpm,Choudhary:2020pxy,Zhang:2022rex,Bamber:2022pbs,Aurrekoetxea:2023jwk,Khalaf:2024nwc,Kim:2025wwj,Roy:2025qaa,LIGOScientific:2025bkz,LIGOScientific:2025ouy} for additional, related work on axion-induced BH SR). 

We show in Fig.~\ref{fig:gagg} our ultimate constraints from GWTC-5 focusing just on the QCD axion (though our analysis applies to other axion-like particles), and compare them to existing SR-based constraints from both the aforementioned X-ray and LVK spin measurements~\cite{Arvanitaki:2014wva,Baryakhtar:2020gao,Witte:2024drg,Ng:2020ruv,Caputo:2025oap, Aswathi:2025nxa}, as well as astrophysical constraints~\cite{AxionLimits} in the context of the coupling $\gagg$ (which is not directly involved in our analysis here). 

\noindent
{\bf Axion Superradiance and Black Hole Spin-Down.---}
\label{sec:superradiance}The formalism of axion-induced SR around BHs is well-established~\cite{Brito:2015oca}: a spinning Kerr BH transfers angular momentum to a bound bosonic field when the field's Compton wavelength is comparable to the BH gravitational radius, leading to an exponential growth of the bound state that depletes the BH mass and spin. Formally, the solution of the Klein-Gordon equation for an axion (or other scalar) field in curved space admits a hydrogen-like separable solution (see, {\it e.g.},~\cite{Brito:2015oca,Dolan:2007mj,Arvanitaki:2010sy,Brito:2017zvb,Baumann:2019eav,Zhu:2020tht,Ellis:2026gwt}); the bound states of the axion cloud can be labeled by quantum numbers $[n, \ell, m]$ (where $m \in [-\ell, \ell]$). For an axion of mass $m_a$ around a BH of mass $M$, the relevant dimensionless coupling is the gravitational fine-structure constant
\begin{equation}
\alpha \equiv G M m_a 
\approx 7.5 \times 10^{-2} 
\left( \frac{m_a}{10^{-12}\,\mathrm{eV}} \right)
\left( \frac{M}{10\, M_\odot} \right) \,.
\label{eq:alpha}
\end{equation}
For $\alpha \ll 1$, the perturbative, hydrogenic solutions have binding energies $\omega_R \approx m_a (1 - \alpha^2 / 2 n_h^2)$ (at $\mathcal{O}(\alpha^2)$), where $n_h = n + \ell + 1$. The SR instability is most efficient for the dominant $\ell = m = 1$, $n = 0$ mode, which we denote simply as $m=1$ in what follows, though the higher $m$ modes probe higher $m_a$, as we discuss further below (for our fiducial analysis we include modes up to $m=6$). Note that the $\ell = m$, $n=0$ states are the fastest-growing modes with that azimuthal number, and so we simply label these modes by $m$. Generally, SR is most effective when $\alpha \sim 0.5$~\cite{Dolan:2007mj}, which implies that, for characteristic BH masses of $M \sim 10 - 100\, M_{\odot}$, SR axion searches are most sensitive to axion masses of order $m_a = 10^{-13} - 10^{-12}$ eV.

\begin{figure}[tb]
\centering
\includegraphics[width=1.0\linewidth]{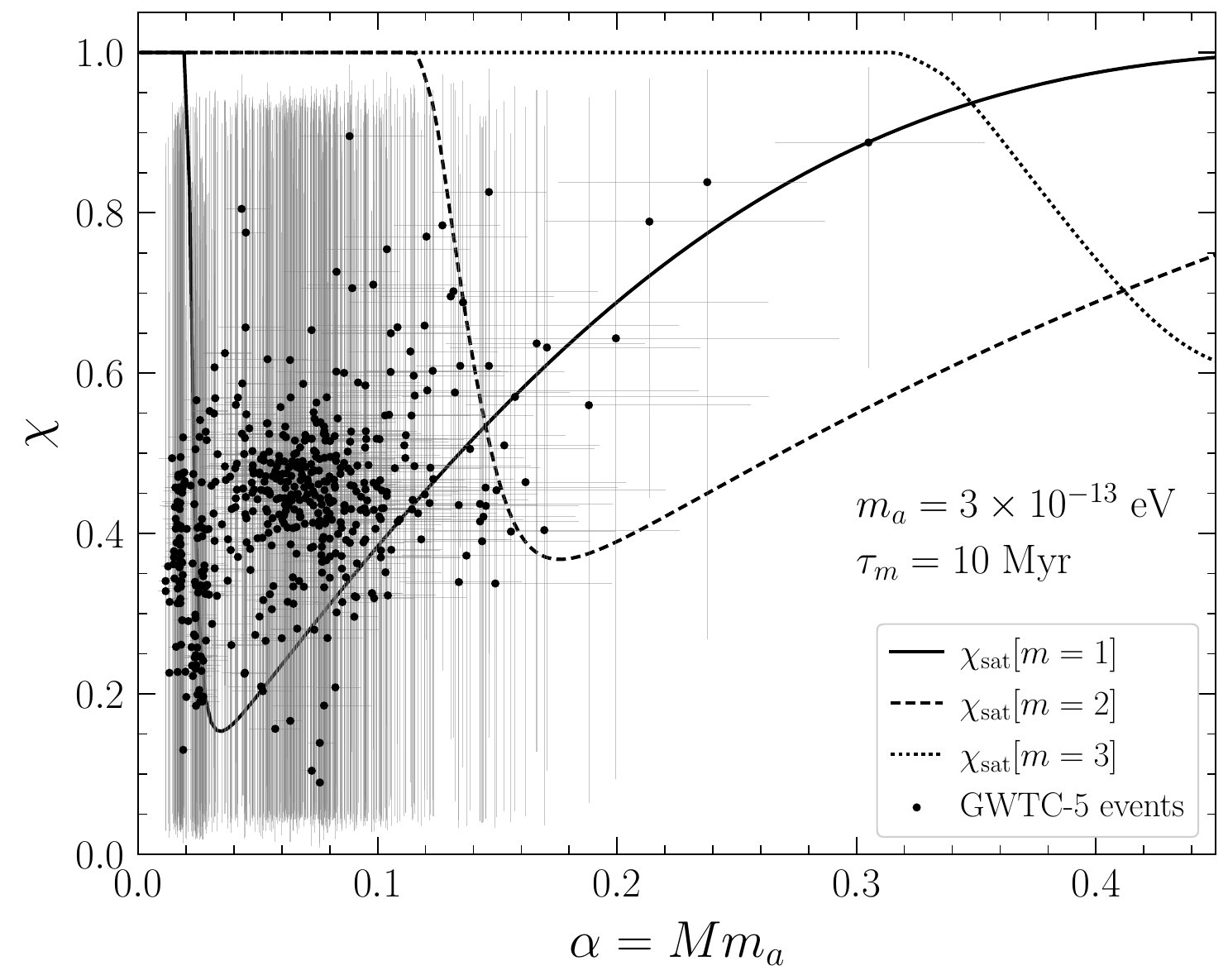}
\caption{The distribution of masses and spins (with error bars reflecting the PE 90\% credible interval) used in our GWTC-5 axion SR search, overlaid with the SR-induced Regge curves (saturated spin trajectories) for the modes $m=1, 2, 3$ at the given axion mass $m_a$ and assuming a merger timescale $\tau_m = 10$ Myr. Under the axion hypothesis, BHs in regions above each Regge curve would be spun down through SR to the saturated spin values along the mode curves.}
\label{fig:regge}
\end{figure}

SR occurs when the bound-state frequency satisfies $\omega_R < m \Omega_H$, with $\Omega_H = \chi/[2 M(1 + \sqrt{1-\chi^2})]$ the angular velocity of the BH horizon and $\chi$ the dimensionless BH spin. The instability extracts angular momentum from the BH, growing the bosonic cloud's occupation number exponentially with $e$-folding time $\tau_{\rm SR} \sim 1/\Gamma_{\rm SR}$. The leading-order rate, computed in, {\it e.g.}, Ref.~\cite{Baumann:2019eav}, scales as $\Gamma_{\rm SR} \propto M^{-1} \alpha^{4 \ell + 5}$ at fixed $\chi$ (see the End Matter (EM)), so heavier BHs and larger couplings spin down more rapidly. A BH born above the critical-spin curve $\chi_c(m_a, M)$, defined as the value of $\chi$ at which $\omega_R = m \Omega_H$, is depleted by the cloud to $\chi \le \chi_c$; below $\chi_c$ the SR condition is no longer satisfied and the instability shuts off. Given the finite lifetimes of astrophysical BHs in binaries, the relevant quantity is the time-bounded critical spin, which we term the saturation spin $\chi_{\rm sat}$.  See Fig.~\ref{fig:regge} for an illustration. The full multi-mode rate formula and the implicit equation for $\chi_{\rm sat}$ are given in the EM.

For SR to act on observationally relevant timescales, the cloud must grow from vacuum fluctuation to macroscopic occupation. The number of $e$-folds required, assuming a typical cloud mass $M_c \sim 0.1 \, M$, is $N_{\rm efolds} \sim \ln(M_c/m_a) \approx 180$ for stellar-mass BHs.  Thus, for axion-induced SR to spin down a BH, the time $\tau_m$ between BH formation and the merger must satisfy $\tau_{\rm grow} < \tau_m$, with $\tau_{\rm grow} \equiv N_{\rm efolds}\tau_{\rm SR}$~\cite{Arvanitaki:2010sy, Arvanitaki:2014wva, Ng:2019jsx}.
The time $\tau_m$ varies across binary evolution channels and from system to system. In our fiducial analysis, we allow each BBH event to have its own $\tau_m$, drawn from a common parametric distribution, with the $\tau_m$ treated as nuisance parameters and marginalized over on a per-event basis. We take as the underlying distribution a log-uniform delay-time distribution (DTD) $p(\tau_m) \propto \tau_m^{-1}$ over $\tau_m \in [10^7, \tau_H]$ yrs, where $\tau_H \sim 1.4\times 10^{10}$ yrs is the Hubble time. This distribution takes into account both  the isolated channel as well as dynamically-assembled binary merger scenarios, and is supported by LVK and long GRB inferences~\cite{Wu:2024znq}. In the Supplementary Materials (SM) we explore the simpler assumption that all binaries share a common $\tau_m$, with a log-uniform prior within the range $\tau_m \in [10^5,10^9]$ yr and find that this choice makes little impact to our final result. (Note that this choice is motivated by Ref.~\cite{Ng:2020ruv}, which fixes $\tau_m = 10^7$ yr for all BBHs.) We also explore the possibility that the DTD is bifurcated into a low- and high-mass population as suggested by, {\it e.g.}, Refs.~\cite{vanSon:2021zpk, Padhyegurjar:2026slt}, though we find that the impact on our final results are minor.

The hydrogen-like rate of Ref.~\cite{Baumann:2018vus} is a leading-order result in $\alpha$, with corrections of $\mathcal{O}(\alpha^2)$ and higher that become non-negligible at $\alpha \gtrsim 0.3$, and which we include in our fiducial analysis (see SM). But at higher $\alpha \gtrsim 0.5$ it is known that the perturbative solution for $\Gamma_{\rm SR}$ becomes an increasingly untenable approximation, whose impact is most deeply felt in higher modes. For the stellar-mass BHs across most of the axion masses $m_a$ for which we are sensitive, $\alpha$ tends to reach values only up to $\alpha \sim 0.5$ and modes less than $m=3$ dominate.
However, as we approach $m_a$ on the high-mass border of our sensitivity, $\alpha$ approaches $\alpha\sim 1$ for some BHs and hence the perturbative solution begins to break down.  As a systematic test, we use the code package released by Ref.~\cite{Witte:2024drg} to numerically solve the SR equations without the small-$\alpha$ approximation, which is important for high $m_a$ and mode number~\cite{Baryakhtar:2020gao}; reassuringly, we find that using the exact SR solutions does not strongly affect our results (see the SM for details).
We also explore in the SM the impact of mixing from tidal effects from the binary companion on cloud growth, finding that for the driving events in GWTC-5 the effects of mixing are likely minor.

\noindent
{\bf Hierarchical Likelihood and GWTC Data.---}\label{sec:methods}We model the natal-spin distribution from which BHs are born as a Beta distribution~\cite{Wysocki:2018mpo,LIGOScientific:2018jsj,LIGOScientific:2020kqk,KAGRA:2021duu}
\begin{equation}
p(\chi \, | \, \alpha, \beta) \propto 
\chi^{\alpha - 1} (1 - \chi)^{\beta - 1} \,, 
\quad \chi \in [0, 1] \,,
\label{eq:beta}
\end{equation}
where $\alpha$ and $\beta$ are population-level hyperparameters. (We consider alternate parameterizations in the SM.) The Beta family is sufficiently flexible to capture natal-spin distributions ranging from those peaked at high spin ($\alpha \gg \beta$) through uniform ($\alpha = \beta = 1$) to those peaked near zero ($\alpha < 1 < \beta$), with the latter regime corresponding to the low-spin natal populations generally favored by current LVK inferences~\cite{KAGRA:2021duu,LIGOScientific:2025pvj}. When including the merger timescales and the axion mass $m_a$, the full set of hyperparameters in our fiducial analysis is $\Lambda = (m_a, \alpha, \beta, \tau_m^1, \tau_m^2, \cdots, \tau_m^{2N})$, with $\tau_m^i$ indexing over the $2N$ BHs in the $N$ BBH mergers. For our fiducial analysis, we assume the uniform priors $\log_{10}(m_a / \mathrm{eV}) \in [-15, -10]$ and $\log_{10}(\alpha), \log_{10}(\beta) \in [-1, 1]$, 
though we characterize the effect of different priors in the SM. In particular, we explore the effect of a non-informative prior on the axion mass $m_a$ using the Jeffreys prior and show that our axion mass exclusion is robust to this prior choice.

The catalog data \mbox{$d = \{d_i\}_{i=1}^{N}$} comprises $N$ BBH events, where each $d_i$ consists of LVK parameter-estimation (PE) posterior samples \mbox{$d_i = \{(M_{1,i}^{(k)}, \chi_{1,i}^{(k)}, M_{2,i}^{(k)}, \chi_{2,i}^{(k)})\}_{k=1}^{K}$}~\cite{KAGRA:2023pio, LIGOScientific:2025snk}, jointly describing the primary and secondary component masses and spins and preserving the per-sample correlations between the two components. We treat each event as a single joint observation of the four-dimensional parameter point $\theta_i = (M_{1,i}, \chi_{1,i}, M_{2,i}, \chi_{2,i})$. Under the axion-signal hypothesis $\mathcal{H}_B$, the joint posterior over the population hyperparameters $\Lambda$ is
\begin{equation}
p(\Lambda \, | \, d, \mathcal{H}_B) \propto \pi(\Lambda) 
\prod_{i=1}^{N} \int d\theta_i \,\, 
p(\theta_i \, | \, \Lambda, \mathcal{H}_B) \, 
p(d_i \, | \, \theta_i) \,,
\label{eq:hyperposterior}
\end{equation}
where $\pi(\Lambda)$ is the hyperprior, $p(d_i | \theta_i)$ is the single-event LVK likelihood, and $p(\theta_i | \Lambda, \mathcal{H}_B)$ is the population prior carrying the SR physics under $\mathcal{H}_B$. Because the PE samples $\{\theta_i^{(k)}\}$ are draws from the single-event posteriors $p(\theta_i | d_i) \propto p(d_i | \theta_i)\, \pi_{\rm PE}(\theta_i)$, the per-event integrals in~\eqref{eq:hyperposterior} are evaluated by importance sampling over the PE samples;
the default LVK PE prior $\pi_{\rm PE}$ is approximately constant on the source-frame parameter space relevant to our analysis, so the importance-sampling reweight is trivial.
The joint $(M_1, \chi_1, M_2, \chi_2)$ correlations within each event thus enter the hierarchical likelihood directly through the joint PE samples, which serve as data-driven priors on the per-event nuisance parameters.
Under $\mathcal{H}_B$ for a given $m_a$, the SR deterministic forward map $\chi_I \to \chi_M(m_a, M, \tau_m, \chi_I)$, with $\chi_I$ the BH's spin at birth, drawn from the Beta distribution, allows us to compute the BH spin $\chi_M$ at merger time.  Note that BHs born above the critical-spin curve $\chi_{\rm sat}(m_a, M, \tau_m)$ are spun down to that curve, while those below preserve their natal spin.
The null hypothesis $\mathcal{H}_A$ (``no axion'') corresponds to removing the spin-down condition entirely and simply sets $\chi_M = \chi_I$.

We apply event-level mass cuts $M_1^{\rm med} \ge 3 \, M_\odot$ and $M_2^{\rm med} \ge 3 \, M_\odot$ to exclude NS candidates~\cite{KAGRA:2021duu, LIGOScientific:2025pvj}, where $M_{1,2}^{\rm med}$ refer to the LVK-posterior medians.
There are $N=257$ BBH merger events in the GWTC-5 catalog after cuts.
The cut criteria are further listed in the SM, as well as a visualization of the top 20 events in GWTC-5 by median spin in the PE distributions. We sample the posterior using the nested sampling algorithm as implemented in \texttt{dynesty}~\cite{2020MNRAS.493.3132S}, with $n_{\rm live} = 2000$ live points and additionally obtain the marginal evidences $Z_B$ and $Z_A$ used to compute Bayes factors.

\begin{figure}[tb]
\centering
\includegraphics[width=1.0\linewidth]{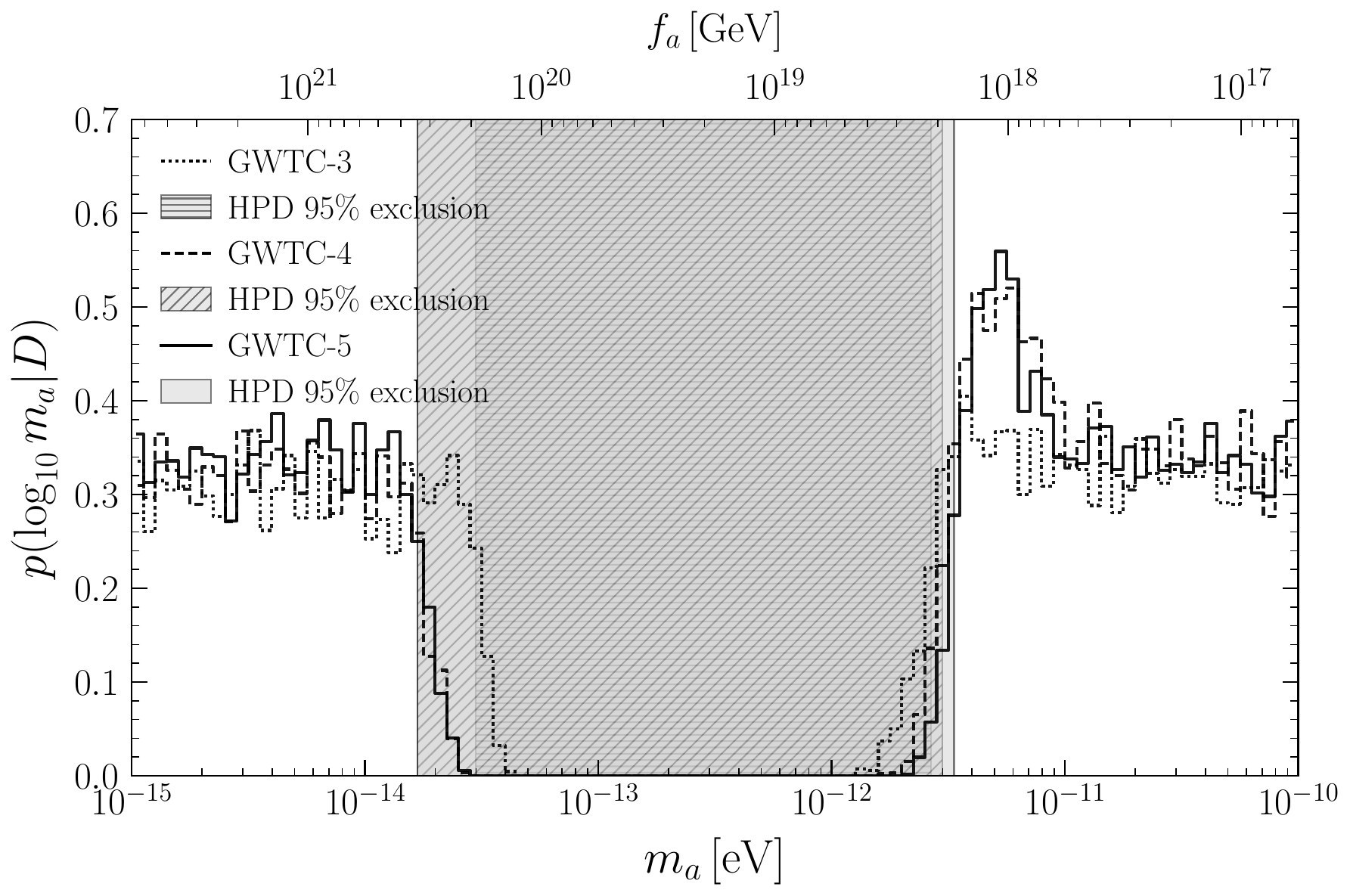}
\caption{The axion mass marginalized posteriors derived from our hierarchical likelihood analysis using BBH data from the GWTC-3, GWTC-4, and GWTC-5 catalogs (note that $f_a$ for the QCD axion is shown as the top x-axis). The posteriors reflect the uniform prior outside the SR region, but display a deep deficit in the SR-active window where axion clouds should have spun down most of the BBH population. 
We additionally illustrate our fiducial 95\% exclusion regions defined through the HPD credible interval, for all three catalog searches.}
\label{fig:posterior}
\end{figure}

\noindent
{\bf Results for GWTC-5.---}\label{sec:results}We present our fiducial search for axion SR in the BBH measurements from the GWTC-5 catalog. The marginal posterior on $\log_{10}(m_a/\mathrm{eV})$ from the GWTC-5 analysis, with the GWTC-3- and GWTC-4-only result overlaid for illustrative purposes, is shown in Fig.~\ref{fig:posterior}. All three posteriors track the log-uniform prior across most of the multi-decade range probed, but exhibit a deep deficit in the SR-active window where SR would have efficiently spun down the bulk of the BBH population on the marginalized $\tau_m$ timescales. The deficit deepens and broadens from GWTC-3 to GWTC-4 to GWTC-5, reflecting the increased constraining power of the larger catalogs. We remark that in all three catalogs there is a conspicuous local maximum in posterior density around $\log_{10}m_a/{\rm eV} \sim -11.4$, which is explained by the fact that at approximately this mass the saturation curves across the modes we consider (with higher modes such as $m=3, 4$ being most important) begin to naturally lie at spin values that sit near the median values of several high-spin GWTC-5 events.
Note that a similar phenomenon was found in Ref.~\cite{Ng:2020ruv}, though that work only operates up to mode $m=3$.

We quantify the exclusion on $m_a$ using the highest posterior density (HPD) credible interval. The smallest region in $\log_{10}(m_a/\mathrm{eV})$ containing $95\%$ of the posterior mass excludes the complement at $95\%$ credibility; for GWTC-5, the excluded region is approximately $\log_{10}(m_a/\mathrm{eV}) \in [-13.8, -11.5 ]$ ($m_a \in [ 1.7 \times 10^{-14}, \, 3.3 \times 10^{-12} ]$ eV), centered on the SR-active window. We also compute the Bayes factor $B^B_A = Z_B / Z_A$ between the axion-signal and null hypotheses.
Integrated over the full $m_a$ prior, $\log B^B_A = 0.22 \pm 0.06$ for GWTC-5, consistent with no evidence for the axion hypothesis~\cite{Kass01061995}.

We verify that the fiducial exclusions are robust to several principal methodological choices, explained further in the SM. First, replacing the log-uniform prior on $m_a$ with the non-informative Jeffreys prior $\pi_J(m_a) \propto \sqrt{I(m_a)}$, where $I(m_a)$ is the Fisher information of the nuisance-marginalized likelihood, estimated over an ensemble of synthetic catalogs, yields an exclusion region of $m_a \in [2.5\times 10^{-14}, \, 2.2 \times 10^{-12}]$~eV, which is comparable to our fiducial results. We also compare our fiducial exclusion to the expected exclusion under the null hypothesis, conducted using Monte Carlo-generated catalog data, and find that our fiducial exclusion region is comparable to the null expectation. We also find our fiducial posteriors and corresponding exclusions are generally robust to astrophysical systematic uncertainties, such as variations on the distribution of $\tau_m$, alternate natal spin parameterizations such as a truncated Gaussian and a Beta-mixture model comprising two weighted Beta functions, and mass-dependent versions of the $\tau_m$ and natal spin distributions.

In addition to the constraint on $m_a$, our analysis simultaneously infers the BH natal-spin population. The posterior medians of the Beta distribution hyperparameters are $\alpha = 0.69^{+0.21}_{-0.12}$ and $\beta = 2.28^{+0.62}_{-0.43}$ for GWTC-5, favoring a low-spin natal population ($\alpha < \beta$), generally consistent with broader LVK population inferences~\cite{KAGRA:2021duu,LIGOScientific:2025pvj,LIGOScientific:2026ctl}.

\noindent
{\bf Discussion.---}In light of evidence that X-ray BH spin measurements for stellar-mass BHs may be biased to high spins, our search here for axion-induced SR in the BBH sample provided in the LVK GWTC-5 catalog represents the strongest lower bound on the mass of the QCD axion as well as the strongest constraints on axion-like particles with masses in the $\sim$$2 \times 10^{-14}$ eV to $3 \times 10^{-12}$ eV mass range. Our results have important implications for terrestrial QCD axion searches, effectively eliminating the lowest-mass part of the QCD axion parameter space targeted by experiments such as CASPEr~\cite{Budker:2013hfa}, as well as implications for string axiverse constructions that predict near log-uniform distributed axion-like particle masses~\cite{Arvanitaki:2009fg}. %

The constraints presented in this work will improve substantially as the LVK network advances over the coming decade. The third portion of the fourth observing run (O4c) concluded in November 2025 and its catalog release is expected in the near term, contributing additional events that will be naturally incorporated into the analysis presented here.  In the fifth observing run, O5, the Advanced LIGO detectors are projected to approximately double the O4 horizon and increase the accessible detection volume by nearly an order of magnitude~\cite{KAGRA:2013rdx}. 
For the present analysis, this translates into three concrete improvements: (i) significantly tighter inference of the natal-spin distribution, breaking degeneracies with the axion-induced spin-down signal; (ii) higher signal-to-noise per-event spin posteriors; and (iii) more high- and low-mass events that extend sensitivity toward lower and higher axion masses. The proposed third-generation detectors Cosmic Explorer~\cite{Reitze:2019iox, Evans:2023euw} and the Einstein Telescope~\cite{Hild:2010id,ET:2019dnz} will, by the 2030s, observe essentially every stellar-mass BBH merger in the observable universe.  We project that this could potentially increase the sensitivity to the QCD axion above  $10^{-11}$ eV, though further work understanding the effects of axion self-interactions is needed to properly account for axion-induced SR at such high QCD axion masses.

\section{Acknowledgments}

{\it
We thank Asimina Arvanitaki, Masha Baryakhtar, Will East, and Nick Rodd for helpful conversations and comments. The authors are supported in part by the DOE award DESC0025293. B.R.S. acknowledges support from the Alfred P. Sloan Foundation.  The work of O.N. is supported in part by the NSF Graduate Research Fellowship Program under Grant DGE2146752. This research used resources of the National Energy Research Scientific Computing Center (NERSC), a U.S. Department of Energy Office of Science User Facility located at Lawrence Berkeley National Laboratory, operated under Contract No. DE-AC02-05CH11231 using NERSC award HEP-ERCAP0023978.   
}

\bibliography{refs}

\clearpage
\appendix

\section{SR Spin-down Rates}
\label{app:superrad}

In this section we provide the explicit rate formula and saturation spin condition for the multi-mode SR instability.

\begin{figure}[tb]
\centering
\includegraphics[width=1.0\linewidth]{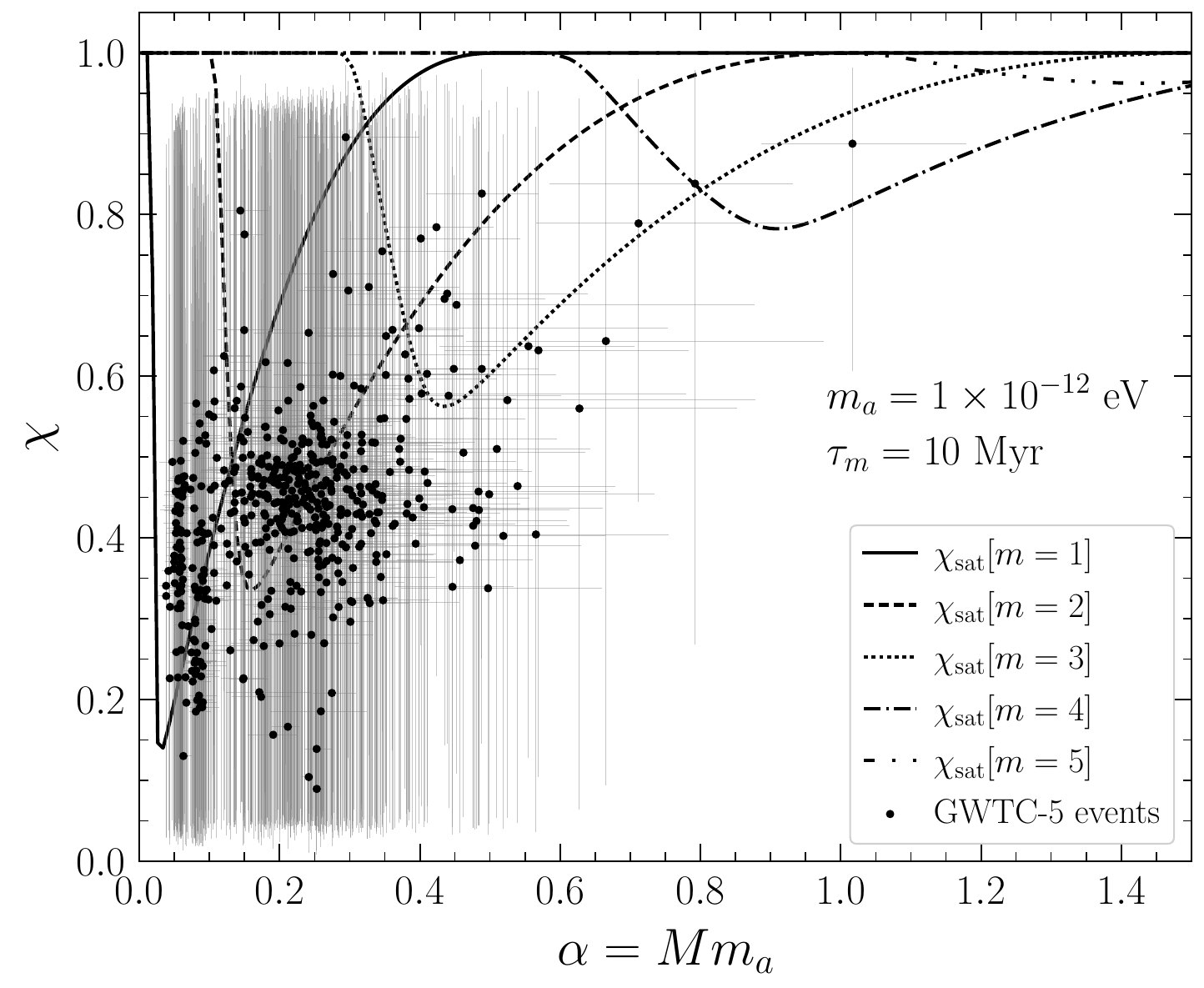}
\caption{The same as Fig.~\ref{fig:regge} but for $m_a = 1 \times 10^{-12}$ eV, for which higher modes such as $m=4$ become more relevant.}
\label{fig:regge_2}
\end{figure}

The leading-order rate of cloud growth for the $[n, \ell,m]$ bound-state mode in the small-$\alpha$ hydrogen-like approximation is given by~\cite{Baumann:2019eav}
\begin{equation}
\Gamma_{\rm SR}^{[n,\ell,m]}(m_a, M, \chi) = 
 \, 2 \alpha^{4\ell + 5} M^{-1} \, 
(m \chi - 2 m_a r_+) \, C_{n\ell m} \, \mathcal{K}(\chi) \,,
\label{eq:gamma_sr}
\end{equation}
where $r_+ = M (1 + \sqrt{1 - \chi^2})$ is the BH outer horizon radius (in units of $M$), $\mathcal{K}(\chi) = \prod_{k=1}^{\ell} [k^2 (1 - \chi^2) + (m \chi - 2 m_a r_+)^2]$, and $C_{n\ell m}$ is a mode-dependent numerical coefficient (with $C_{0,1,1} = 1/24$ for the dominant $\ell = m = 1$, $n = 0$ mode~\cite{Detweiler:1980uk_1} ). The rate scales as $\Gamma_{\rm SR} \propto M^{-1} \alpha^{4\ell + 5}$ for fixed $\chi$, confirming that, for most of the $\alpha$ we consider, the dominant growth occurs in the lowest $\ell = m$ modes.  The higher $m$ states become important at large axion masses, see Fig.~\ref{fig:regge_2}.

\begin{figure*}[!htb]
\centering
\begin{minipage}{0.49\linewidth}
\centering
\includegraphics[width=\linewidth]{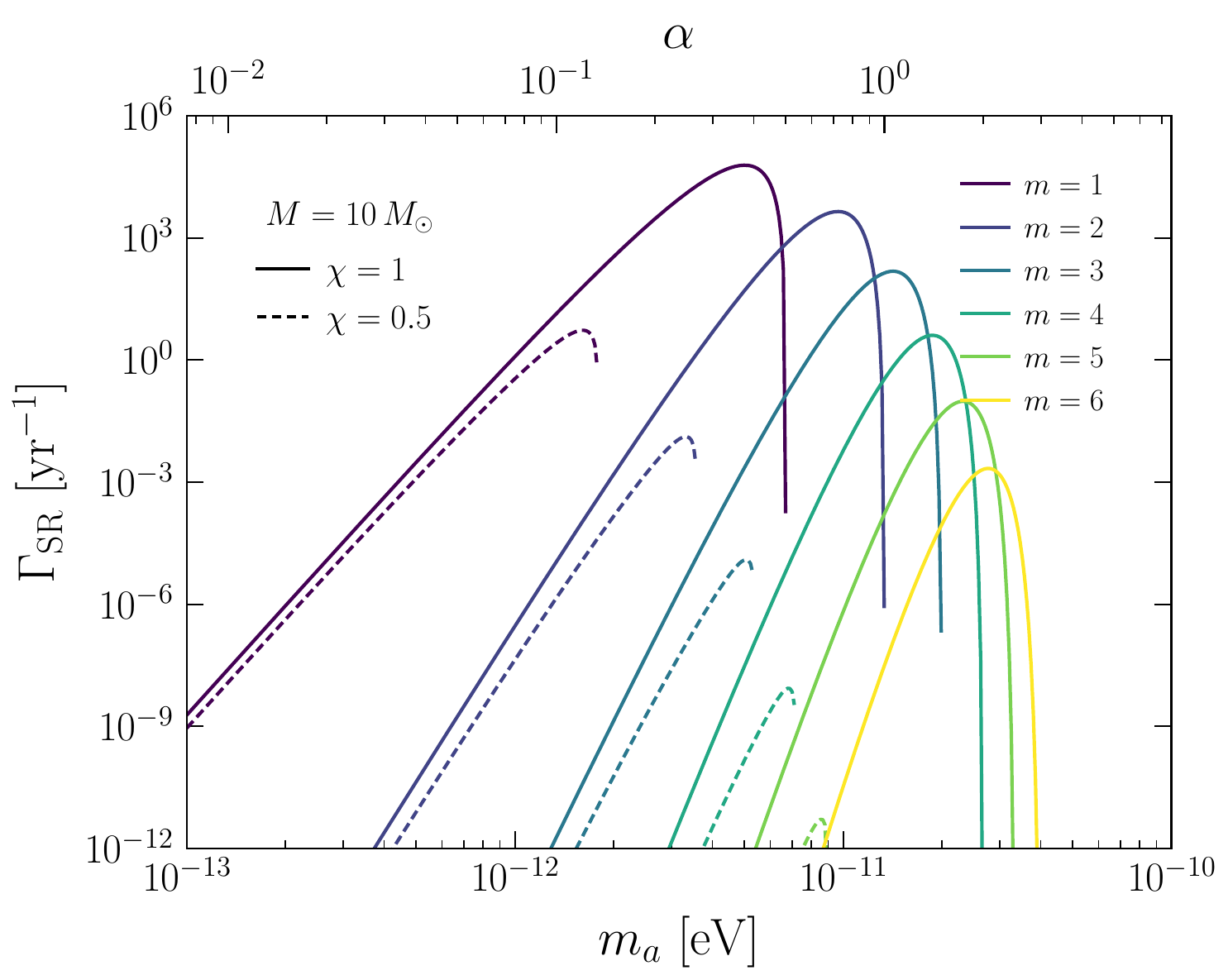}
\end{minipage}\hfill
\begin{minipage}{0.49\linewidth}
\centering
\includegraphics[width=\linewidth]{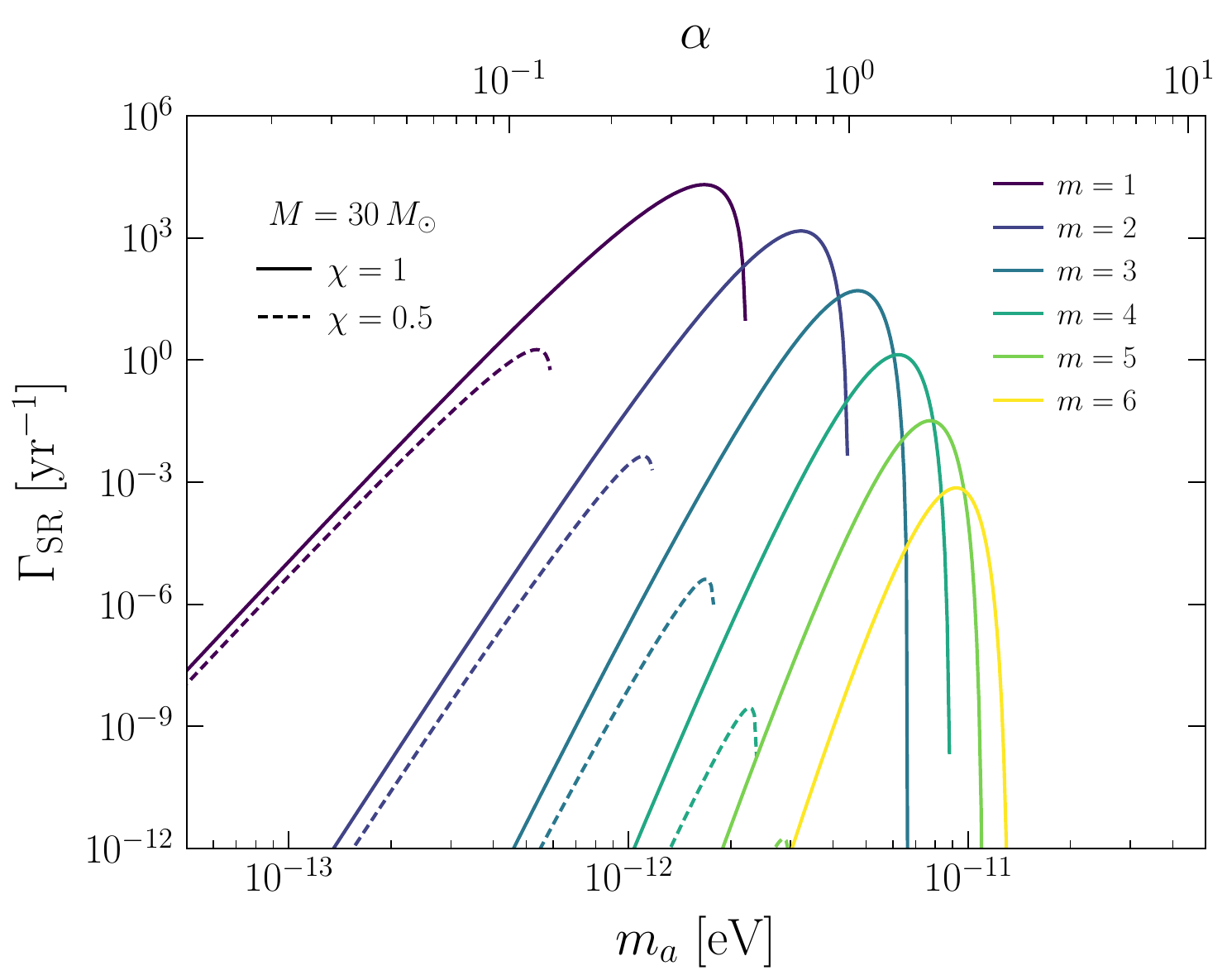}
\end{minipage}

\vspace{0.1cm}

\begin{minipage}{0.49\linewidth}
\centering
\includegraphics[width=\linewidth]{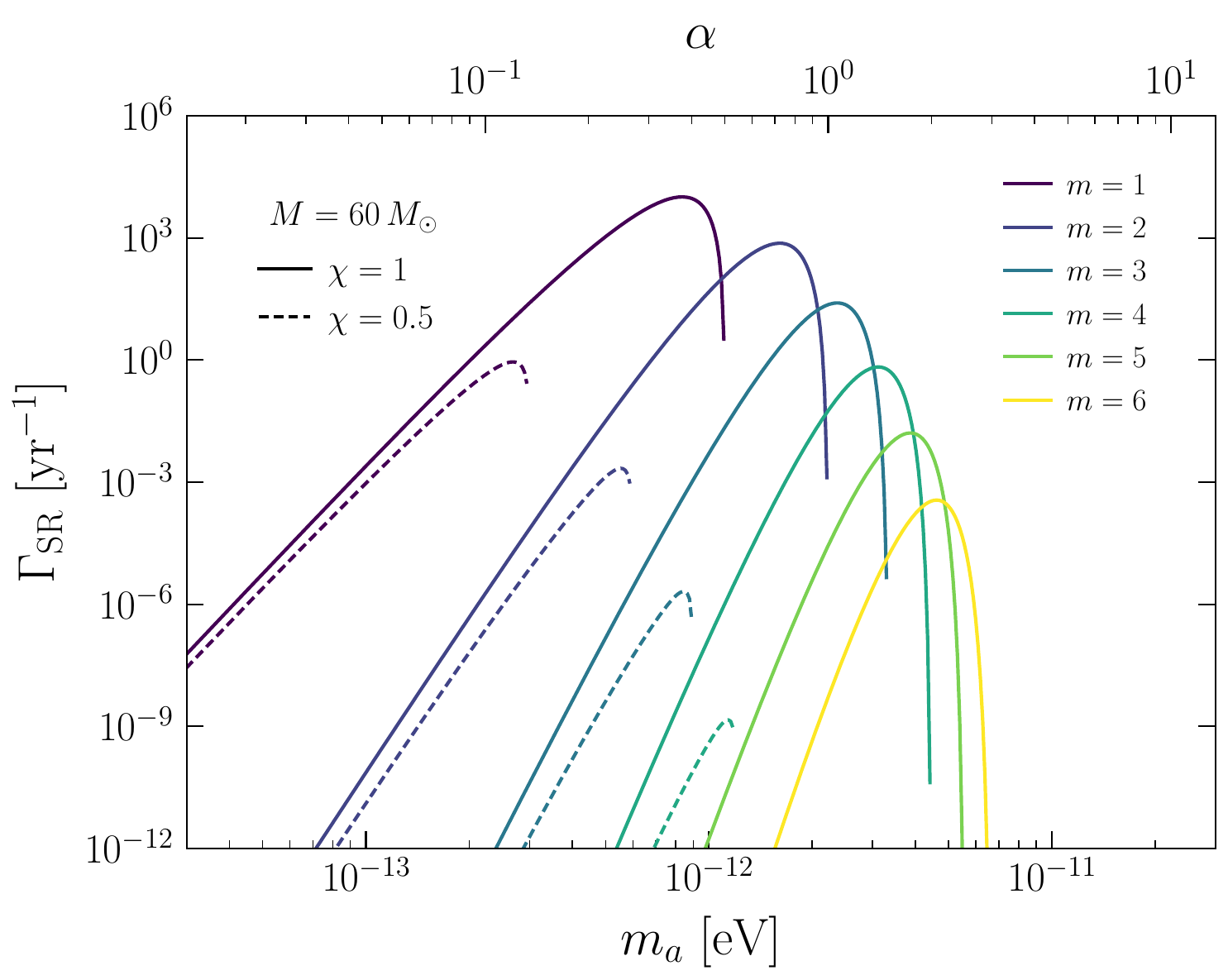}
\end{minipage}\hfill
\begin{minipage}{0.49\linewidth}
\centering
\includegraphics[width=\linewidth]{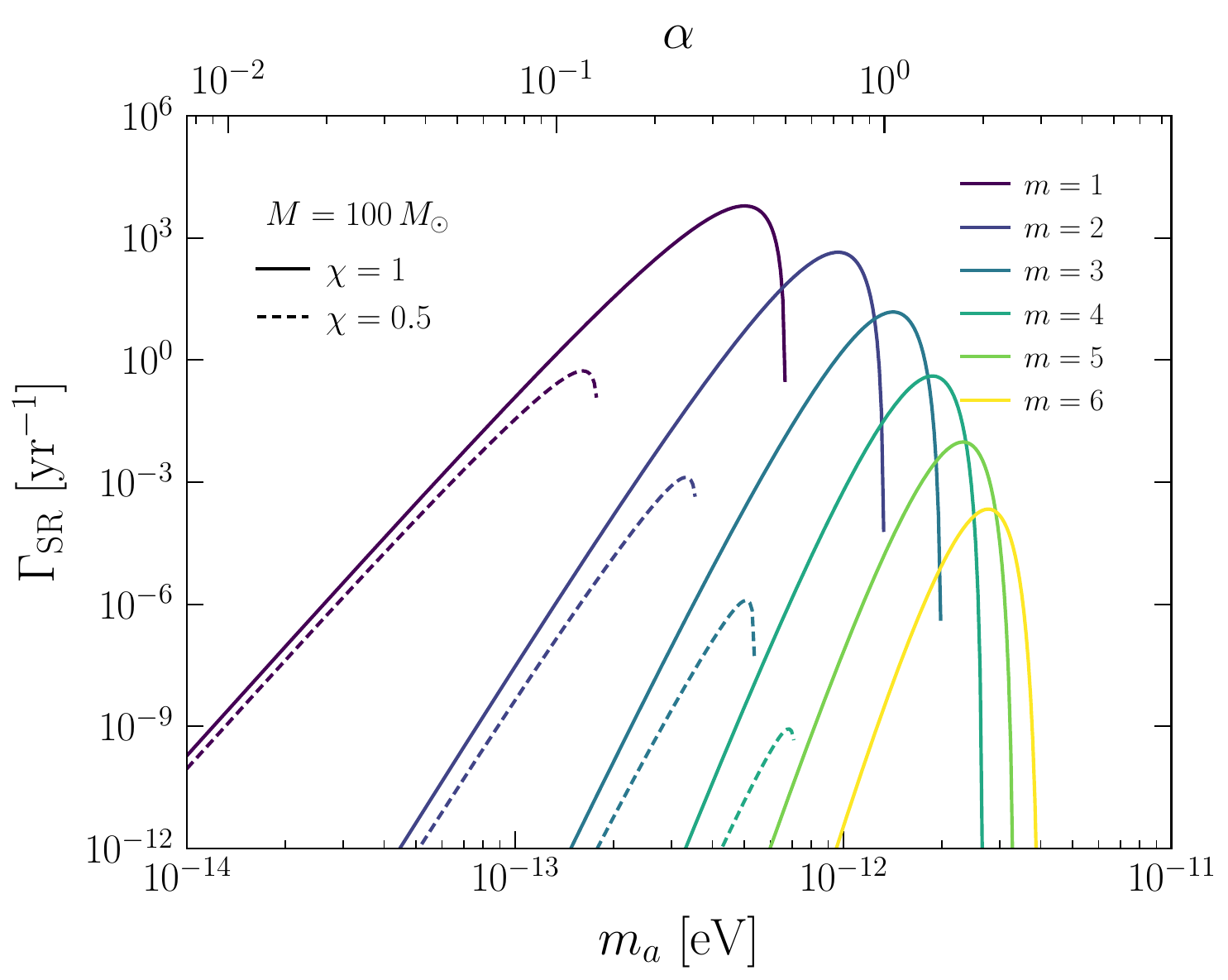}
\end{minipage}
\caption{The superradiance cloud growth rate $\Gamma_{\rm SR}$ for $\chi = 1$ (solid) and $\chi=0.5$ (dashed), as a function of axion mass $m_a$ (equivalently, $\alpha$, shown as the top x-axis), shown for modes 1 through 6. Each panel represents a given BH mass, $10$$M_{\odot}$ (top left), $30$$M_{\odot}$ (top right), $60$$M_{\odot}$ (bottom left), $100$$M_{\odot}$ (bottom right).}
\label{fig:sr_rates}
\end{figure*}

We include the $m=1$ through $m=6$ modes in our fiducial analysis, which covers the parameter space in $(\chi, \alpha)$ spanned by GWTC-5 events across sensitive $m_a$ values. We illustrate the SR rates for modes 1 through 6 for four representative BH masses $M$ across $m_a$ (equivalently $\alpha$) for two different spins $\chi$ in Fig.~\ref{fig:sr_rates}, remarking that generally higher modes, higher spins, and smaller BH masses allow SR to occur for larger axion masses $m_a$, with generally higher SR rates.

For SR to occur, the condition $\Gamma_{\rm SR}(m_a, M, \chi) > 0$ (equivalently $\omega_R < m \Omega_H$), implies (up to $\mathcal{O}(\alpha^2)$ corrections), the spin to satisfy
\begin{equation}
\frac{\chi}{2 \, (1 + \sqrt{1 - \chi^2})} = 
\frac{\alpha}{m} \,,
\label{eq:chic}
\end{equation}
where we note that this implies $\alpha$ is constrained to follow $\alpha < m/2$. Further, for SR to occur within the given BH merger timescale $\tau_m$, we solve for the time-bounded critical spin, which we term the saturated spin $\chi_{\rm sat}$,  and which is the solution to the equation
\begin{equation}
\Gamma_{\rm SR}^{[n, l, m]} (m_a, M, \chi_{\rm sat}) = \frac{N_{\rm efolds}}{\tau_{m}}
\label{eq:chi_sat}
\end{equation}
so long as the mode has time to grow, i.e. $\tau_m > \tau_{\rm grow} \equiv N_{\rm efolds} \tau_{\rm SR}^{[n, l, m]}(\chi_I)$ where $\chi_I$ is the spin at the onset of SR. We note that a given BH can have multiple modes excited within $\tau_m$ (so long as $\tau_m > \tau_{\rm grow}$), where it successively spins down to the saturated spin of the highest mode achievable within $\tau_m$. 

\clearpage
\onecolumngrid

\begin{center}
  \textbf{\large Supplementary Material for No Evidence for Superradiant Axions in LIGO-Virgo-KAGRA GWTC-5 Binary Black Hole Spins}\\[.2cm]
  \vspace{0.05in}
  {Orion Ning, Benjamin R. Safdi, and Catherine Welch}
\end{center}

\twocolumngrid

\setcounter{equation}{0}
\setcounter{figure}{0}
\setcounter{table}{0}
\setcounter{section}{0}
\setcounter{page}{1}
\makeatletter
\renewcommand{\theequation}{S\arabic{equation}}
\renewcommand{\thefigure}{S\arabic{figure}}
\renewcommand{\thetable}{S\arabic{table}}

\onecolumngrid

This Supplementary Material (SM) includes further details and descriptions regarding the GWTC catalogs used and the hierarchical likelihood analysis, as well as additional tests such as the Jeffreys prior, expectations under the null hypothesis, and the results of systematic variations around our fiducial analysis.

\section{GWTC Catalog Details}
\label{app:gwtc}

\begin{figure}[!h]
\centering
\includegraphics[width=0.75\textwidth]{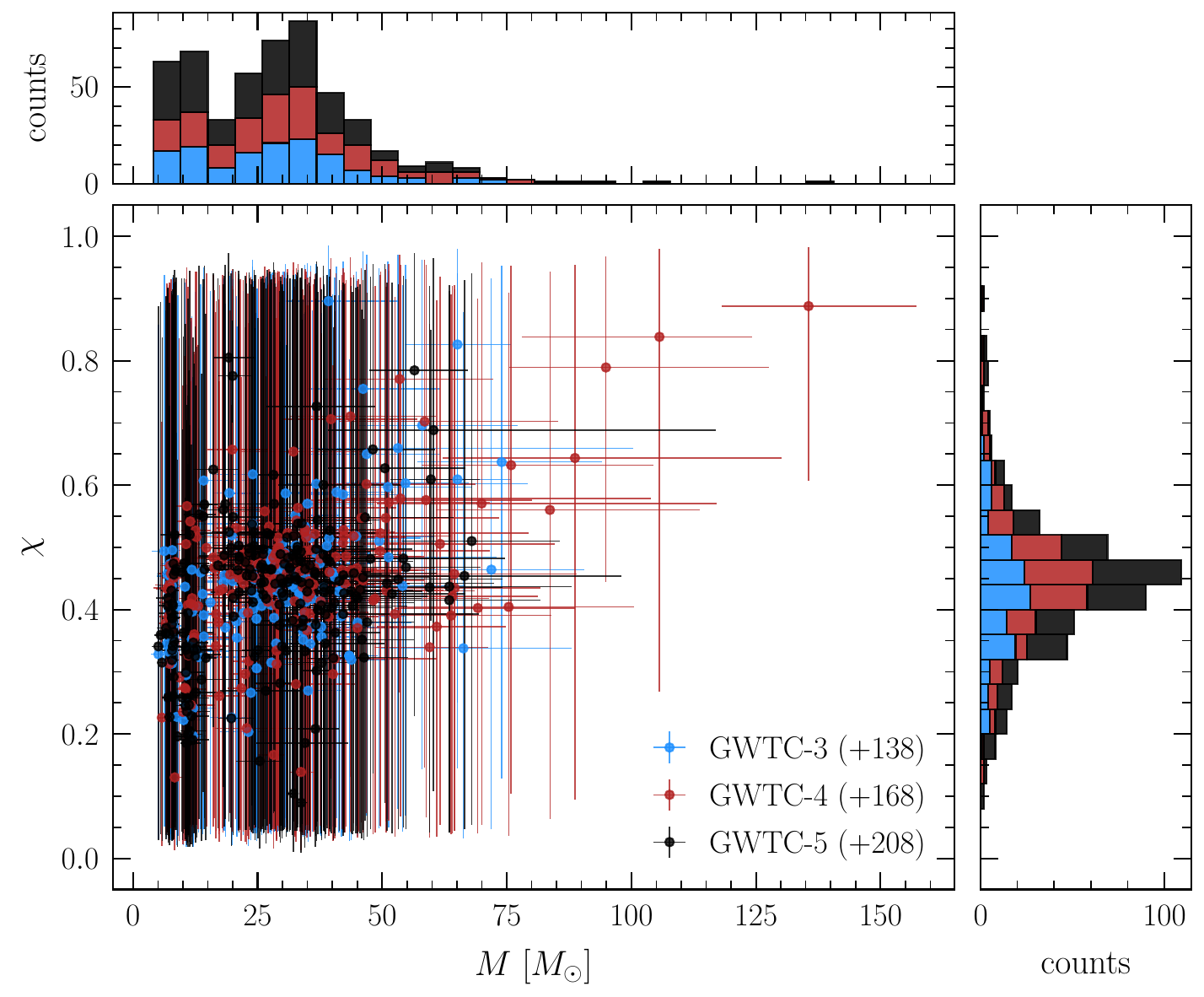}
\caption{The total GWTC-5 ($N=257$) BBH event distribution in the mass-spin plane $(M, \chi)$, for both primary and secondary BHs, used in our fiducial analysis. Each point shows the PE median with $90\%$ credible intervals. Markers are colored by catalog additions, with the number in the legend indicating the number of events belonging to each catalog addition which pass our fiducial threshold event cuts. The top and right subplots are histograms indicating the distribution of (median) masses and spins.}
\label{fig:catalog_scatter}
\end{figure}

\begin{figure}[!h]
\centering
\includegraphics[width=\textwidth]{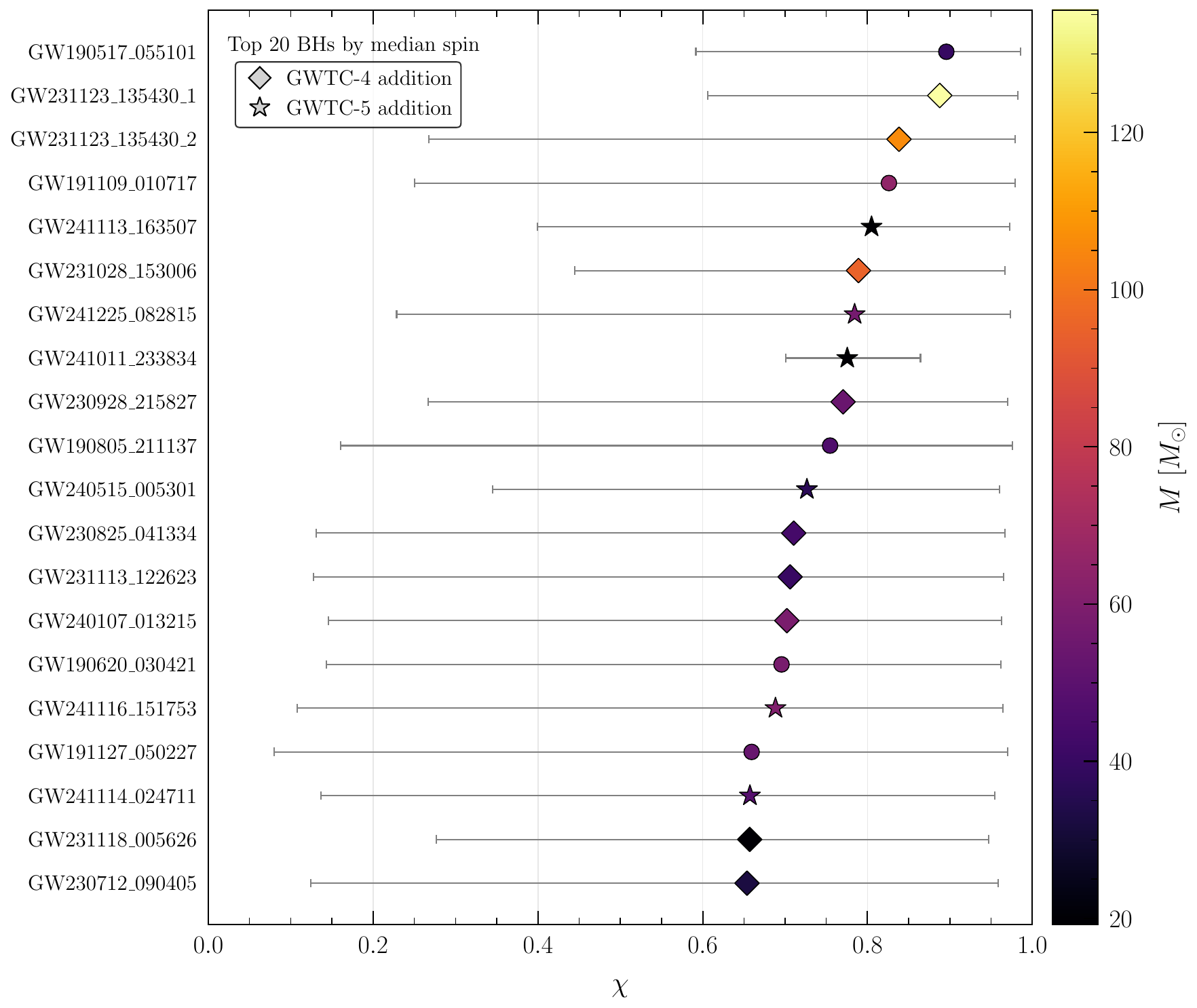}
\caption{The top 20 BHs in the GWTC-5 catalog by median spin, $\chi$. Errorbars indicate the 90\% credible intervals on the spin, and the colorbar indicates the mass $M$. All events are primary BHs with the exception of GW231123\_135430, whose primary and secondary BHs both qualify within the top 20. We also show with a diamond marker events which were newly added in GWTC-4, and a star marker for events which were newly added in GWTC-5.}
\label{fig:top20}
\end{figure}

We use the BBH event lists from the LVK Collaboration data releases for GWTC-3, GWTC-4, and GWTC-5~\cite{KAGRA:2023pio, LIGOScientific:2025snk, LIGOScientific:2026wfs}, restricting to events with false-alarm rate $\mathrm{FAR} < 1~\mathrm{yr}^{-1}$ and identified as BBH candidates. We additionally apply event-level lower mass cuts on the median PE-inferred component masses, $M_1^{\rm med} \geq 3 \, M_\odot$ and $M_2^{\rm med} \geq 3 \, M_\odot $ to exclude NS and ambiguous compact-object candidates~\cite{KAGRA:2021duu, LIGOScientific:2025pvj}. We do not apply an upper mass cut, in contrast to the $M_1 < 75 \, M_\odot$ cut adopted in the GWTC-2 analysis of Ref.~\cite{Ng:2020ruv}, since several high-mass systems have been found in GWTC-4 (and even individually analyzed in the context of axion SR~\cite{Caputo:2025oap, Aswathi:2025nxa}). After cuts, the GWTC-3 catalog contains $N = 69$ BBHs, the GWTC-4 catalog contains $N = 153$ BBHs, and the GWTC-5 catalog contains $N=257$ BBHs. The full GWTC-5 event distribution in the mass-spin plane is shown in Fig.~\ref{fig:catalog_scatter}, with each event plotted at its PE median with $90\%$ credible intervals. The coloring indicates to which GWTC addition each event belongs, and the top and right subplots are histograms over (median) masses and spins, respectively. We show in Fig.~\ref{fig:top20} the top 20 BBH events in GWTC-5, by median spin, which are generally the most constraining events in the SR analysis. In that figure, we also indicate the 90\% credible intervals on $\chi$, as well as the associated BH masses and whether an event was added as part of GWTC-4 or GWTC-5.

We apply the mass cuts only at the event level, not at the level per PE-sample. Per-sample cuts would preferentially remove the tails of each event's mass PE distribution, precisely at the edges of the SR active region, and would therefore introduce a bias in the inferred SR signal. A small fraction ($\sim 0.7\%$) of GWTC-5 PE samples extends to masses above $M_1 = 100 \, M_\odot$ with the most extreme case being GW231123\_135430 (median $M_1 \approx 135 \, M_\odot$). Our PEs for GWTC-3, GWTC-4, and GWTC-5 are drawn from the public LVK PE data releases~\cite{KAGRA:2023pio, LIGOScientific:2025snk, LIGOScientific:2026wfs}, using the \texttt{Mixed} sample dataset across waveform families. For events added in GWTC-5, we follow the same conventions as previous catalogs and construct the \texttt{Mixed} dataset consistently (the GWTC-5 release~\cite{LIGOScientific:2026wfs} at the time of this work did not release an official \texttt{Mixed} dataset waveform for all events).

The original GWTC-2 analysis of Ref.~\cite{Ng:2020ruv} excluded possible NSNS and NSBH candidates via a lower primary mass cut of $5 \, M_{\odot}$; again, in our analysis we impose a $M_i^{\rm med} \ge 3 \, M_\odot$ cut for both $M_1$ and $M_2$. Beyond the explicit lower and upper mass cuts, our GWTC-3 catalog shares the $45$ events used in Ref.~\cite{Ng:2020ruv}, with the remainder being events analyzed with updated PEs in GWTC-3. The additional BBHs provided by GWTC-4 and GWTC-5 increase the total number of analyzed BBHs by over five times compared to that used in the GWTC-2 analysis, with many new BBHs detected in these later catalogs possessing high spins important for driving SR sensitivity (see Fig.~\ref{fig:top20}).

\section{Hierarchical Likelihood}
\label{app:likelihood}

In this section we  provide further details about the construction of our hierarchical likelihood, whose overall posterior is given in Eq.~\eqref{eq:hyperposterior}, and whose construction is similar to that used in Refs.~\cite{Ng:2019jsx, Ng:2020ruv}. In particular, we focus on the term $p(\theta_i \, | \,\Lambda, \, \mathcal{H}_B)$, which is responsible for implementing the SR physics into the overall likelihood. The SR forward model is a deterministic piecewise map from the natal spin $\chi_I$ to the merger spin $\chi_M$ at fixed $(m_a, M, \tau_m)$. In the following discussion we will describe the SR as occurring for one active mode $m$, but the generalization to multiple modes (for which we implement $m=1$ through $m=6$) is straightforward. The BH preserves its natal spin ($\chi_M = \chi_I$) when $\chi_I$ is below the time-bounded critical spin, which we term the saturated spin $\chi_{\rm sat}(m_a, M, \tau_m)$; otherwise, under SR the BH spins down to that value when $\chi_I$ lies in the natal-spin band that triggers mode-$m$ growth. Pushing the Beta natal-spin distribution through this map gives a two-branch decomposition of the population spin probability under the axion-signal hypothesis:
\begin{equation}
p(\chi_M | \Lambda, \mathcal{H}_B) = 
\underbrace{p(\chi_M | \alpha, \beta) \, \Theta(\chi_{\rm sat} - \chi_M)}_{\text{no-SR (``$\mathcal{L}_{\rm OFF}$'')}} + \underbrace{f_{\rm SR} \, \delta(\chi_M - \chi_{\rm sat})}_{\text{saturation (``$\mathcal{L}_{\rm SAT}$'')}} \,,
\label{eq:pushforward}
\end{equation}
where $\Theta$ is the Heaviside step function and the saturation-branch weights $f_{\rm SR}(\alpha, \beta) = \int_{\chi_{\rm sat}}^1 p(\chi_I | \alpha, \beta) \, d\chi_I$ give the Beta-CDF probability mass over each mode's natal-spin trigger band. The per-event likelihood correspondingly factorizes as $\mathcal{L}_i^B = \mathcal{L}_i^{\rm OFF} + \mathcal{L}_i^{\rm SAT}$, with
\begin{align}\mathcal{L}_i^{\rm OFF} &= \frac{1}{K} \sum_{k=1}^{K} p(\chi_i^{(k)} | \alpha, \beta) \, \mathbb{I}\!\left[\chi_i^{(k)} < \chi_{\rm sat}(m_a, M_i^{(k)}, \tau_m)\right] \,,\label{eq:Loff}\\[3pt]
\mathcal{L}_i^{\rm SAT} &=  f_{\rm SR}(\alpha, \beta) \, \widetilde{p}_i\!\left(\chi_{\rm sat}(m_a, M_i)\right) \,,\label{eq:Lsat}
\end{align}
where $\widetilde{p}_i$ is the single-event LVK posterior density on the merger spin. As discussed in the main text, we evaluate the full integrated likelihoods via importance sampling; for each event we use $K = 2000$ PE samples. We sample primary and secondary mass and spin posteriors using a common-index thinning, in which the $k$-th primary PE sample $(M_1^{(k)}, \chi_1^{(k)})$ is paired with the $k$-th secondary PE sample $(M_2^{(k)}, \chi_2^{(k)})$ from the same posterior chain, preserving the joint mass-spin correlations within each event. This joint-spin population model would better encapsulate formation-channel effects shared between the two component masses in realistic BBH systems. Within the importance sampling we adopt the mass prior assumptions in Ref.~\cite{Ng:2019jsx, Ng:2020ruv} where the canonical Salpeter power law $\pi(M_1) \propto 1/M_1^{2.35}$ defines the primary mass prior, while uniform draws of $q \equiv M_2/M_1$ between 0 and 1 (where $M_2 \leq M_1$) define the secondary mass prior $\pi(M_2)$.

\section{Jeffreys Prior Test}
\label{app:jeffreys}
In our fiducial analysis we assume a log-uniform prior on the axion mass $m_a$. To check the robustness of the exclusion under a non-informative prior choice, we repeat our fiducial analysis under the Jeffreys prior~\cite{jeffreys:1946}, which is invariant under reparameterizations of $m_a$ and is by construction the ``data-only'' prior in the sense that it weights each value of $m_a$ by the Fisher information that the likelihood encodes about it.

For a multi-parameter model with hyperparameters $\Lambda$ and likelihood $\mathcal{L}(\Lambda)$, the Jeffreys prior on a single parameter $\theta = \log_{10} m_a$, marginalized over the remaining nuisance hyperparameters $\nu = (\alpha, \beta, \tau_m^i)$, is
\begin{equation}
\pi_J(\theta) \propto \sqrt{I(\theta)} \,, 
\quad I(\theta) = \mathbb{E} 
\!\left[ \left( \frac{\partial \log \mathcal{L}(\theta \,|\, D')}{\partial \theta} \right)^2 \right] \,,
\label{eq:jeffreys}
\end{equation}
where the expectation is taken and averaged over synthetic Monte Carlo catalogs $D'$ generated for each $m_a$ under the axion hypothesis $\mathcal{H_B}$, while the nuisance hyperparameters are fixed by the null (a description for the Monte Carlo procedure is detailed in the next section). Effectively, the Jeffreys prior naturally encodes which regions of $m_a$ are most informative with respect to the likelihood for typical realizations of observed data.

\begin{figure}[htb]
\centering
\includegraphics[width=0.49\textwidth]{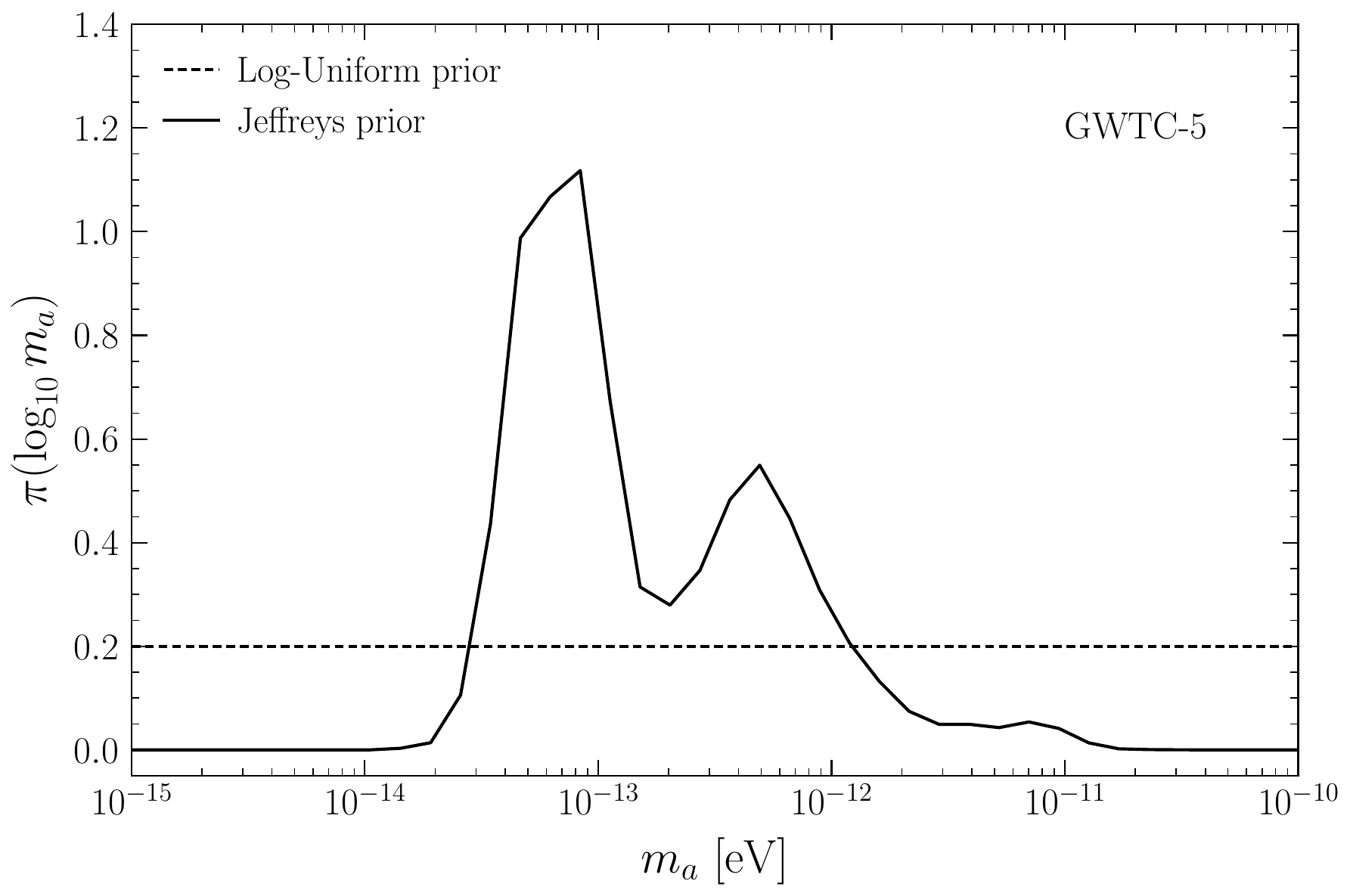}
\includegraphics[width=0.49\textwidth]{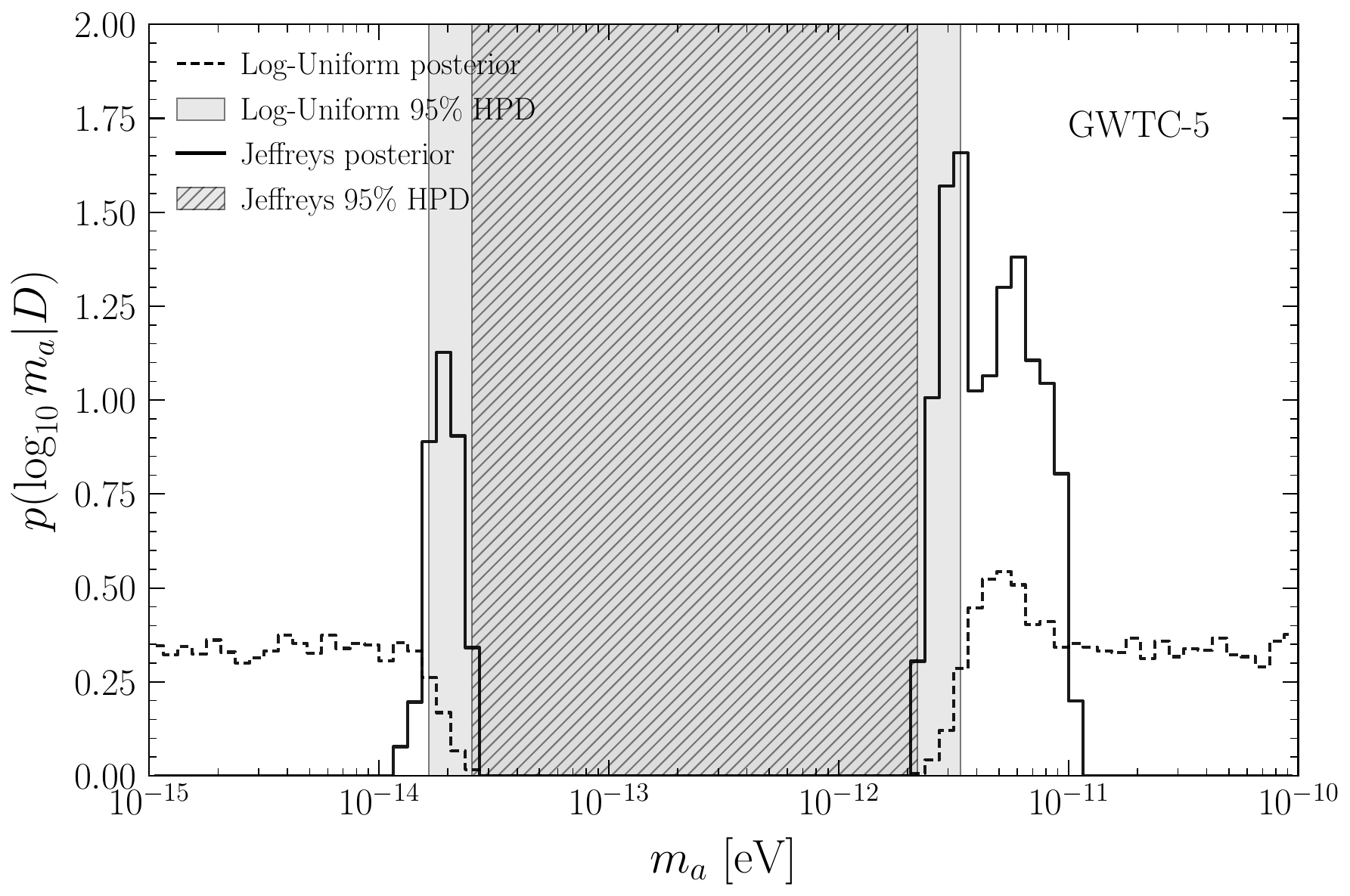}
\vspace{-0.2cm}
\caption{Left: The Jeffreys prior $\pi_J(\log_{10}m_a)$ for GWTC-5. Right: The resulting GWTC-5 posteriors under Jeffreys (solid) and log-uniform (dashed) priors, with 95\% HPD exclusion regions shaded. Note that the Jeffreys prior is non-zero only in regions where the data carry positive Fisher information about $m_a$. The two posteriors yield largely consistent exclusion regions, with the Jeffreys exclusion region being slightly more conservative.}
\label{fig:jeffreys}
\end{figure}

We compute $\pi_J(\theta)$ over $\log_{10}m_a \in [-15, -10]$ with the resulting prior shown in the left panel of Fig.~\ref{fig:jeffreys}, where it broadly has support within the edges of the SR-active regions, and is identically zero elsewhere. The Jeffreys-reweighted posterior on $\log_{10}m_a$, with all other conditions the same as in our fiducial run, is shown in the right panel of Fig.~\ref{fig:jeffreys} alongside the fiducial log-uniform posterior, with the $95\%$ HPD exclusion regions of both shaded for comparison. Both posteriors yield relatively consistent exclusion regions over the canonical SR-sensitive window. The $95\%$ HPD interval under the Jeffreys prior is $m_a \in [2.5\times 10^{-14}, \, 2.2 \times 10^{-12}]$~eV, broadly overlapping (but slightly more conservative than) the fiducial log-uniform exclusion of $m_a \in [1.7\times 10^{-14}, \, 3.3 \times 10^{-12}]$~eV.

\section{Expectations Under the Null}
\label{app:expected}

The exclusion presented in our fiducial analysis is derived from the actual LVK BBH catalog data. To characterize the exclusion the analysis would produce on typical realizations of the null hypothesis $\mathcal{H}_A$, that is, the expected depletion of the $m_a$ posterior in the SR-active window in the absence of any axion signal, we perform a Monte Carlo analysis over synthetic null catalogs and run the fiducial pipeline on each, providing a baseline comparison to our fiducial exclusion using the LVK data.

For each of $S=20$ random seeds, we generate a synthetic catalog matching the size of the real GWTC-5 catalog. For each synthetic event $i$, we draw the component masses $(M_{1,i}, M_{2,i})$ by sampling from the empirical distribution of GWTC-5 PE-median masses, subject to the same mass cuts as the real-data analysis. Natal spins for both components are drawn independently from the population Beta distribution at fixed hyperparameters $(\alpha_{\rm null}, \beta_{\rm null})$ inferred from the real-data fit under the null hypothesis. Under $\mathcal{H}_A$ no SR acts, so the observed spins equal the natal spins. We synthesize PE samples per event for each component by drawing from a split-normal distribution centered on the true $(M, \chi)$ values, with asymmetric widths $(\sigma^-, \sigma^+)$ sampled from the empirical distribution of per-event PE uncertainties in the real GWTC-5 catalog. This mimics the realistic LVK PE-measurement broadening on a per-event basis.

\begin{figure}[htb]
\centering
\includegraphics[width=0.49\textwidth]{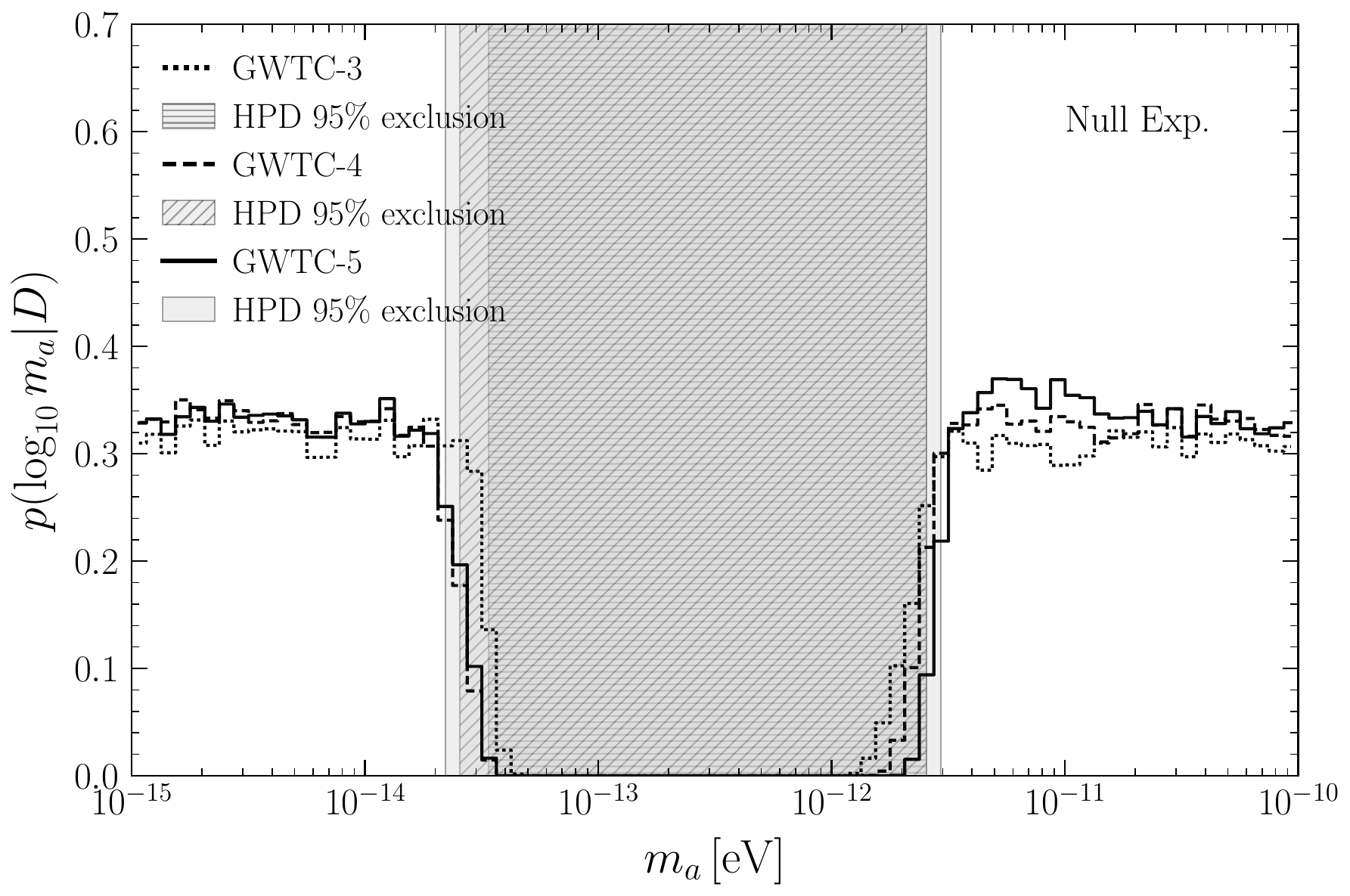}
\includegraphics[width=0.49\textwidth]{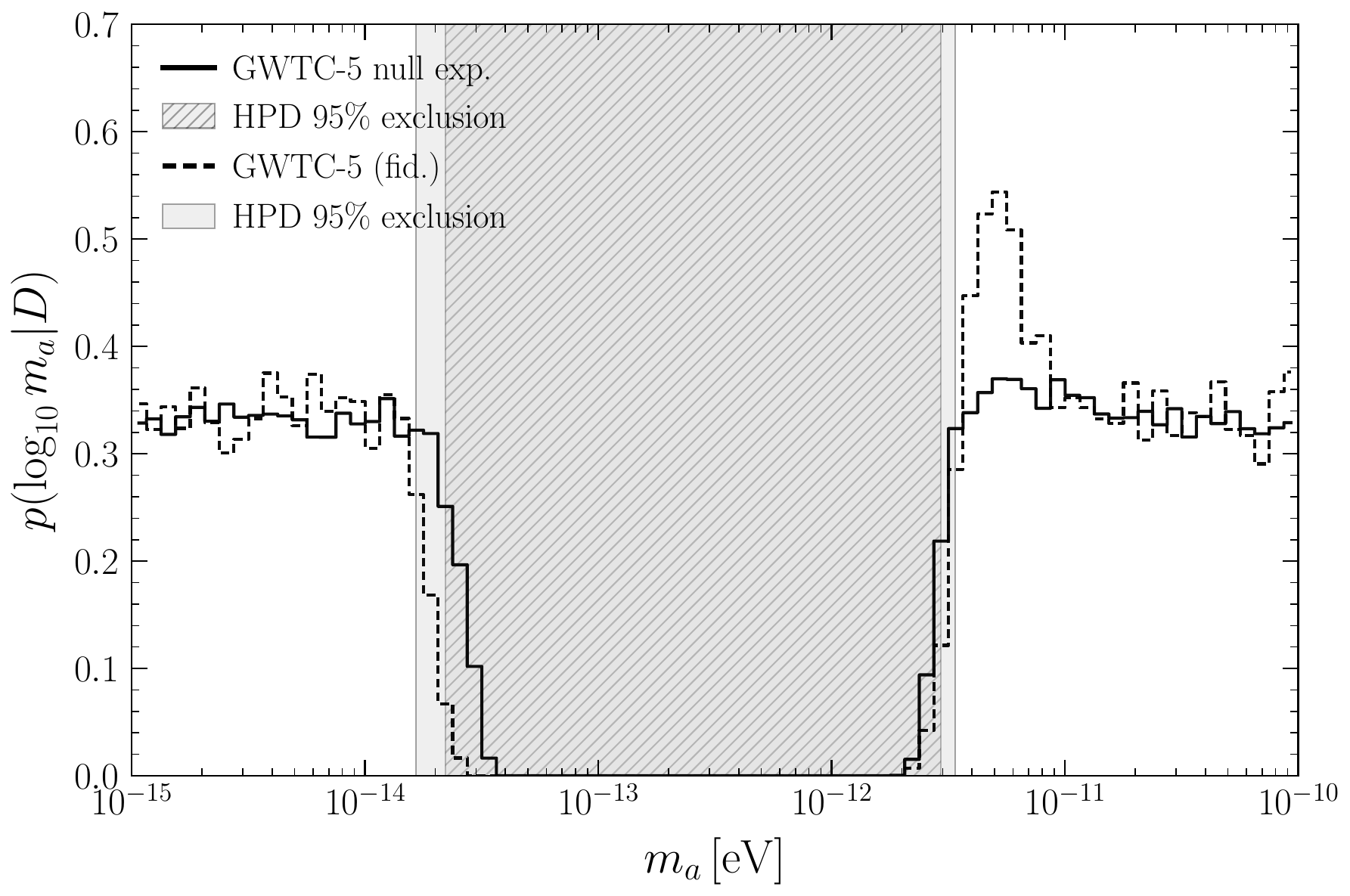}
\vspace{-0.2cm}
\caption{Left: The expected $m_a$ posteriors and 95\% exclusion regions under the null hypothesis, following the Monte Carlo procedure described in the main text, across our three catalogs. Right: The expected posteriors and exclusion regions for GWTC-5 compared to our fiducial analysis using real GWTC-5 data.}
\label{fig:null}
\end{figure}

We then run our fiducial nested-sampling analysis on each synthetic catalog, marginalizing over all hyperparameters $\Lambda = (m_a, \alpha, \beta, \tau_m^i)$ under the same framework as the real-data analysis. Pooling and averaging the per-seed $m_a$ posteriors gives the expected null posteriors shown in Fig.~\ref{fig:null}. The left figure shows the expectations under the null hypothesis across our three main GWTC catalogs, while the right figure compares our GWTC-5 expected posterior with our fiducial posterior, finding similar 95\% exclusion agreement with the observed real-data exclusion. The real data excludes slightly more on the low-mass end of the exclusion region due to the presence of high-mass moderately-high-spin systems in the real data, and on the high-mass end the real data builds up a small peak due to the abundance of observed BBHs with moderate spins naturally sitting close to SR saturation curves.

\section{Signal Injection Test}
\label{app:signal_inj}

In this section we perform signal injection tests on the GWTC-5 catalog to assess the detectability of an axion signal. We do this by constructing synthetic catalogs with an injected SR signal at a known axion mass $m_a^{\rm inj}$ and natal-spin population, and analyze them with the same hierarchical likelihood as the real-data analysis. These synthetic catalogs are generated using the same procedure as the Monte Carlo techniques used to derive expectations under the null hypothesis, with one modification: the observed spin of each BH component is now the post-SR spin obtained by applying the SR forward map at the injected axion mass $m_a^{\rm inj}$. Operationally, natal spins below the saturation spin curve $\chi_{\rm sat}(m_a^{\rm inj}, M, \tau_m)$ are preserved at their natal values; those above are mapped to the corresponding saturation spin $\chi_{\rm sat}$ of the highest mode that triggered within $\tau_m$. Synthetic PE samples are then generated around the post-SR spins with empirically-sampled widths, as in the Monte Carlo technique described earlier for the null hypothesis.

\begin{figure}[htb]
\centering
\includegraphics[width=0.49\textwidth]{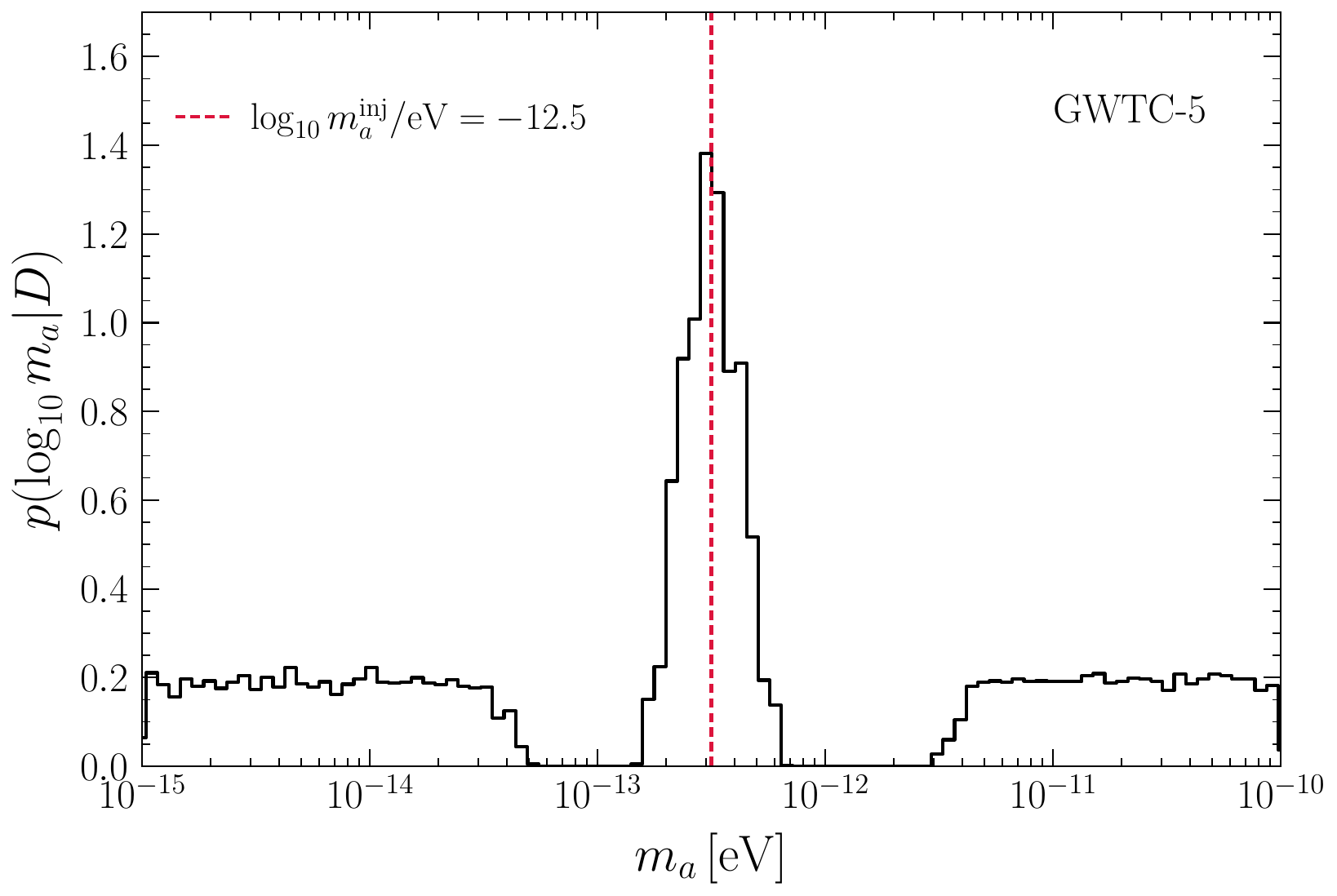}
\includegraphics[width=0.49\textwidth]{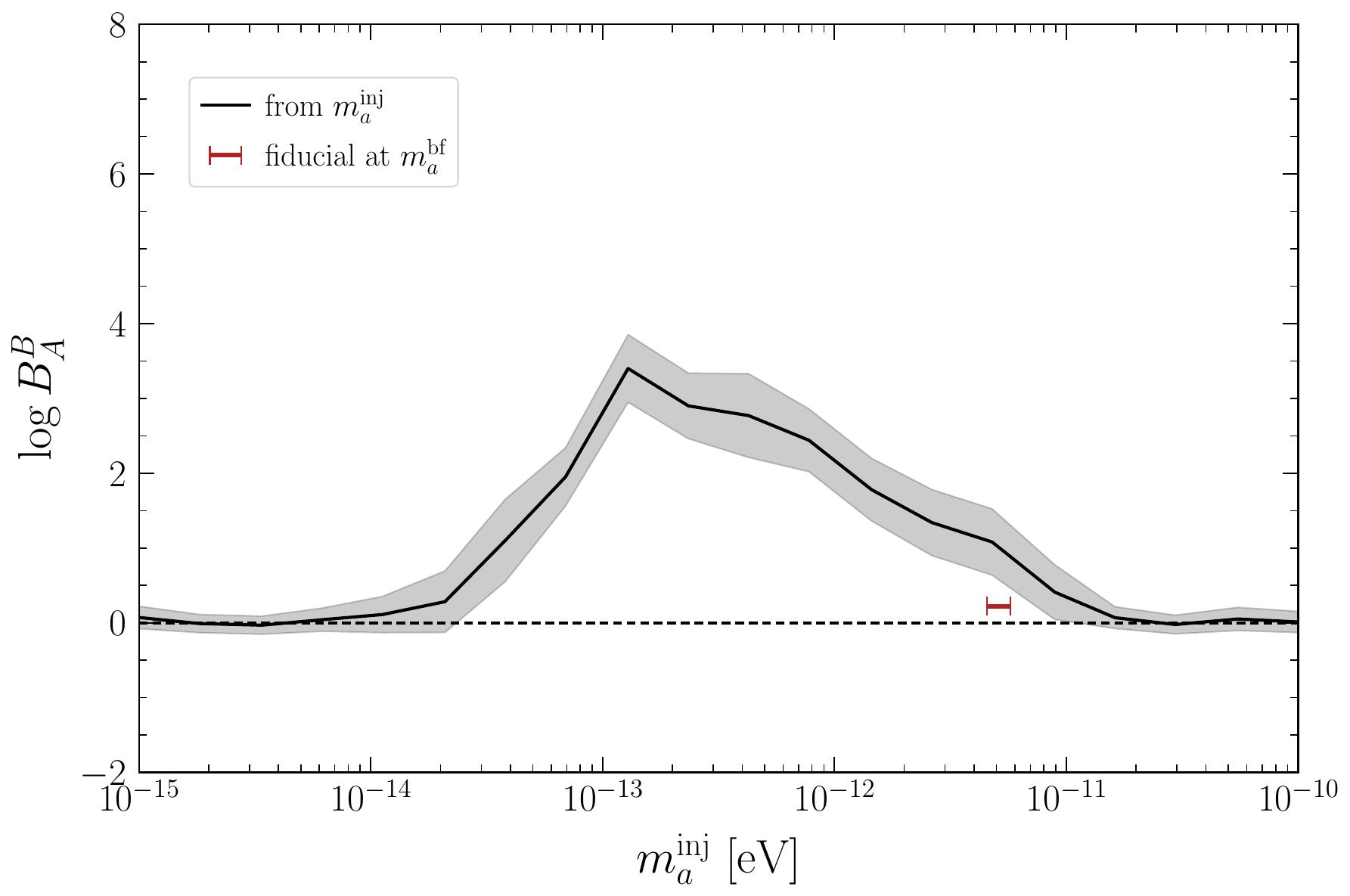}
\caption{(Left) The resulting posterior from an injected axion signal at $m_a^{\rm inj} \approx 3 \times 10^{-13}$~eV (dashed red) derived from synthetic Monte Carlo catalogs for GWTC-5, demonstrating the visible posterior peak recovered around the injected mass value. (Right) An illustration of the log Bayes factors, along with their $1\sigma$ containment, for multiple Monte Carlo-based signal injection tests across our axion mass prior. The red horizontal bar denotes the $1\sigma$ containment interval for the best-fit mass from the observed data.}
\label{fig:siginj}
\end{figure}

We then run our fiducial nested-sampling analysis on the injected catalog across different axion masses $m_a$, demonstrating this explicitly at an injected axion mass in the center of our $m_a$ prior range, $\log_{10}m_a^{\rm inj}/{\rm eV} = -12.5$ or $m_a^{\rm inj} \approx 3 \times 10^{-13}$ eV, in the left panel of Fig.~\ref{fig:siginj}. Here we see our posterior visibly peaked around the injected value, with a strong Bayes factor of $\log B^B_A = 2.8 \pm 0.5$, showing that our pipeline can demonstrate evidence for the SR signal when one is present in the data. More broadly, we show our distribution of Bayes factors from signal injections over various $m_a^{\rm inj}$ in the right panel of Fig.~\ref{fig:siginj} with the bands indicating $1\sigma$ containment intervals from Monte Carlo realizations. In that figure, we also indicate the Bayes factor and $1\sigma$ containment interval for the best-fit mass from the actual GWTC-5 observed data used in our fiducial analysis.

\section{Superradiance for higher $\alpha$}
\label{app:superrad_more}

\begin{figure}[htb]
\centering
\includegraphics[width=0.5\textwidth]{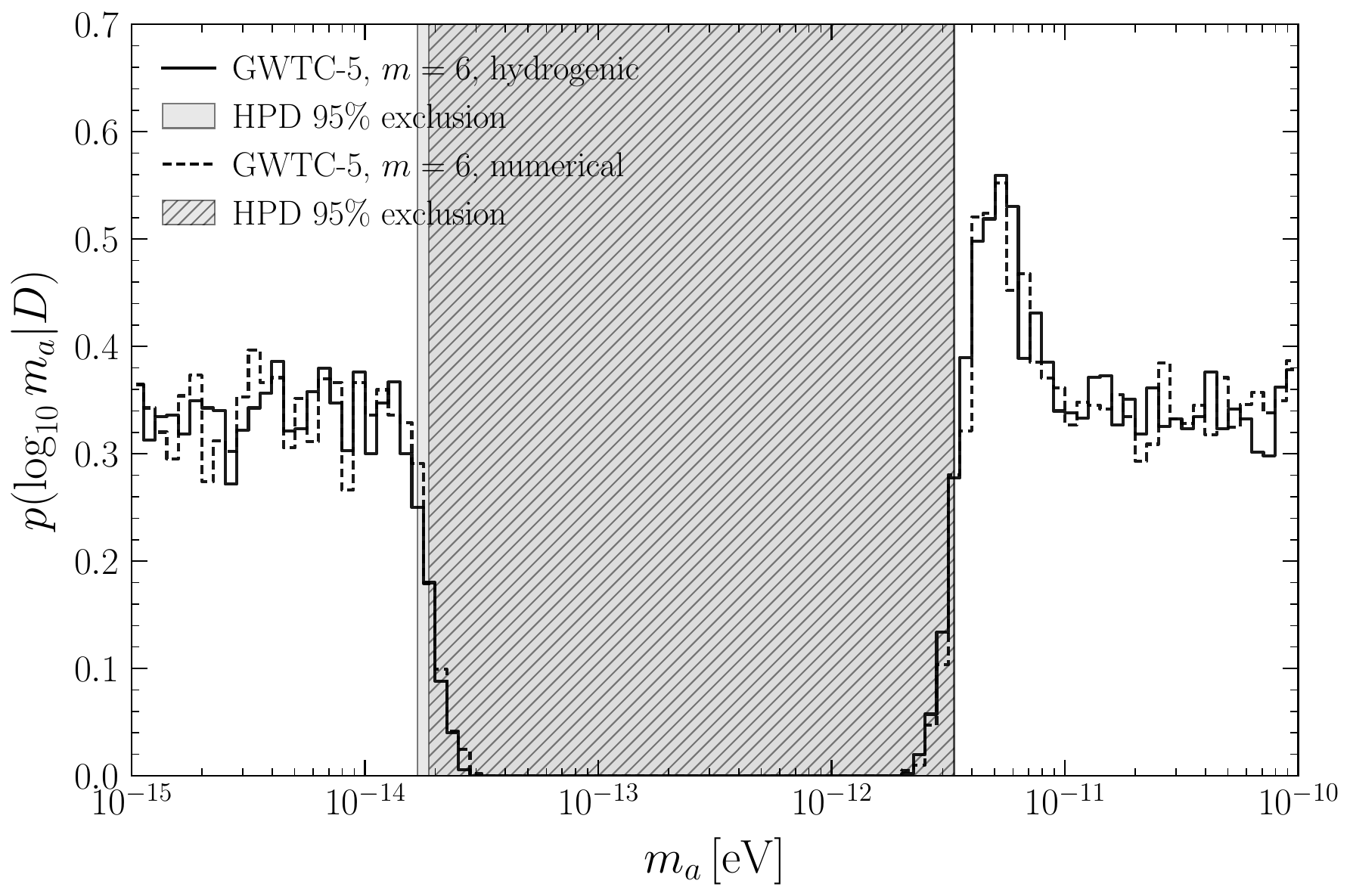}
\caption{A comparison of our fiducial axion mass $m_a$ marginalized posterior from GWTC-5, and the equivalent posterior assuming the SR rates calculated from the codebase in Ref.~\cite{Witte:2024drg}. In the former case, we use hydrogenic analytic formulae to compute $\Gamma_{\rm SR}$ which is formally valid for small $\alpha$, and in the latter case numerical techniques are used to compute $\Gamma_{\rm SR}$ for arbitrary $\alpha$. Both cases are computed up to mode $m=6$, and the resulting posteriors suggest that for our fiducial GWTC-5 catalog search the effects of going beyond the small $\alpha$ approximation are minor.} 
\label{fig:compare_num_alpha}
\end{figure}

The hydrogen-like approximation underlying Eqs.~\eqref{eq:gamma_sr}--\eqref{eq:chic} is exact in the $\alpha \to 0$ limit; corrections enter at $\mathcal{O}(\alpha^2)$ and become numerically significant at $\alpha \gtrsim 0.5$. We retain the leading $\alpha^2$ correction in the bound-state frequency $\omega_R$ throughout our fiducial analysis, which captures the dominant fine-structure-like shift of the critical-spin curve at moderate $\alpha$. While we note that the impact on the inferred axion-mass exclusion is bounded by the contribution of high-$\alpha$ PE samples to the analysis, which is small since the bulk of the catalog probes $\alpha \lesssim 0.5$ for sensitive $m_a$ masses, we conduct an additional systematic test where we numerically calculate $\Gamma_{\rm SR}$ \textit{without} assuming the small $\alpha$ hydrogenic approximation, using the codebase in Ref.~\cite{Witte:2024drg}, which uses numerical Leaver techniques to calculate $\Gamma_{\rm SR}$ for arbitrary $\alpha$ and arbitrary mode $m$. We illustrate the results of our axion search in GWTC-5 using this codebase, going up to $m=6$, in Fig.~\ref{fig:compare_num_alpha}, in which we find similar posteriors and exclusion regions as under our fiducial configuration. This justifies the retainment of our fiducial hydrogenic formulation of SR, which has the advantage of being more computationally efficient. Nonetheless, generally one should be cautious, as higher modes are more susceptible to effects such as self-interactions; though when including self-interactions with the same code base, we find they have a generally negligible effect on the final spins when assuming the QCD axion decay constant, which we focus on in this work. While the BH’s final spin $\chi$ may be altered by up to $\sim$0.05 in some cases, self-interactions in this regime do not prevent modes that build up in the linear theory from doing so, though a more intensive analysis for other $f_a$ values could be an avenue for future work.

\section{Astrophysical Systematic Tests}
\label{app:astro}
\begin{figure}[htb]
\centering
\includegraphics[width=0.49\textwidth]{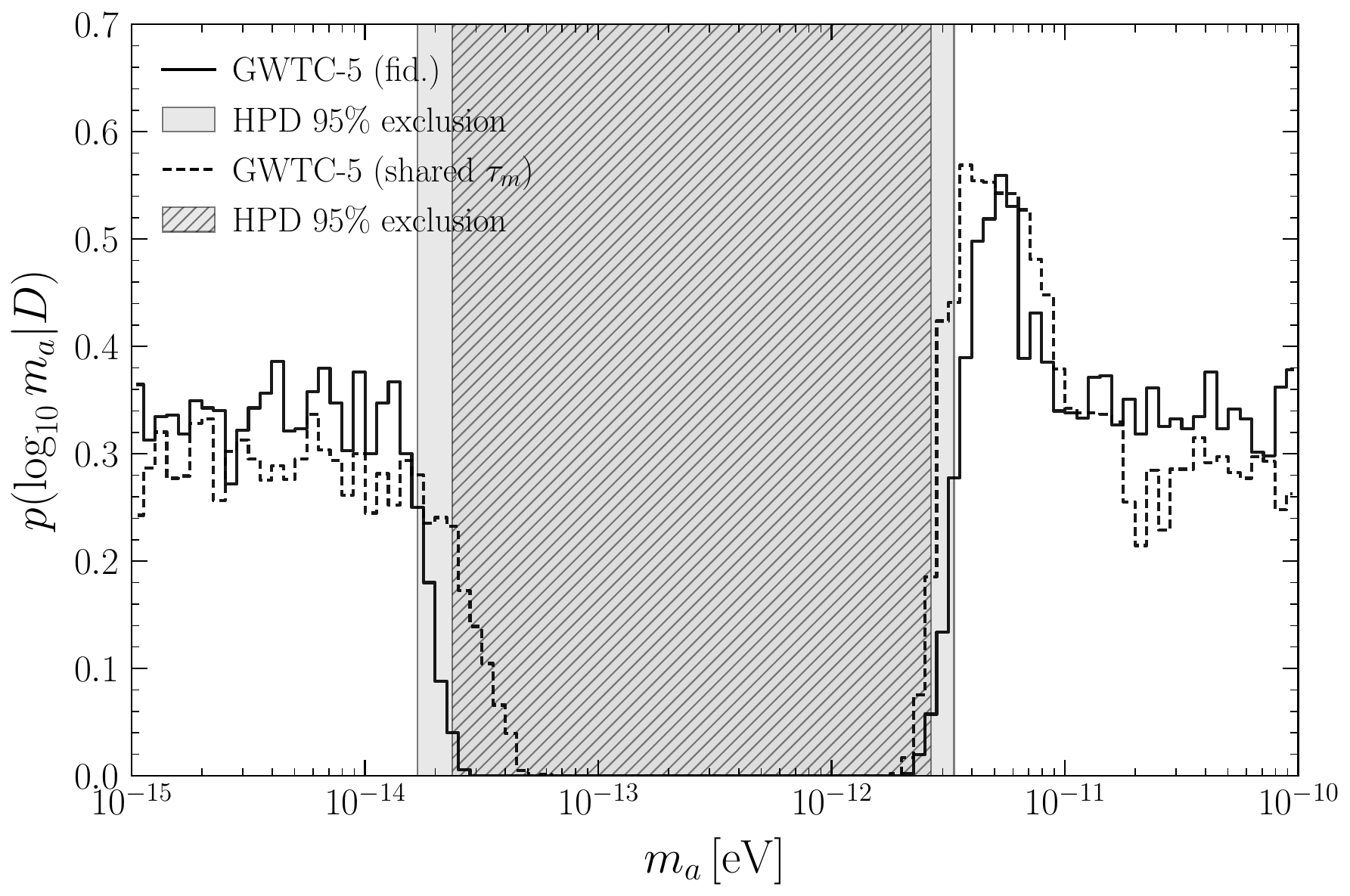}
\includegraphics[width=0.49\textwidth]{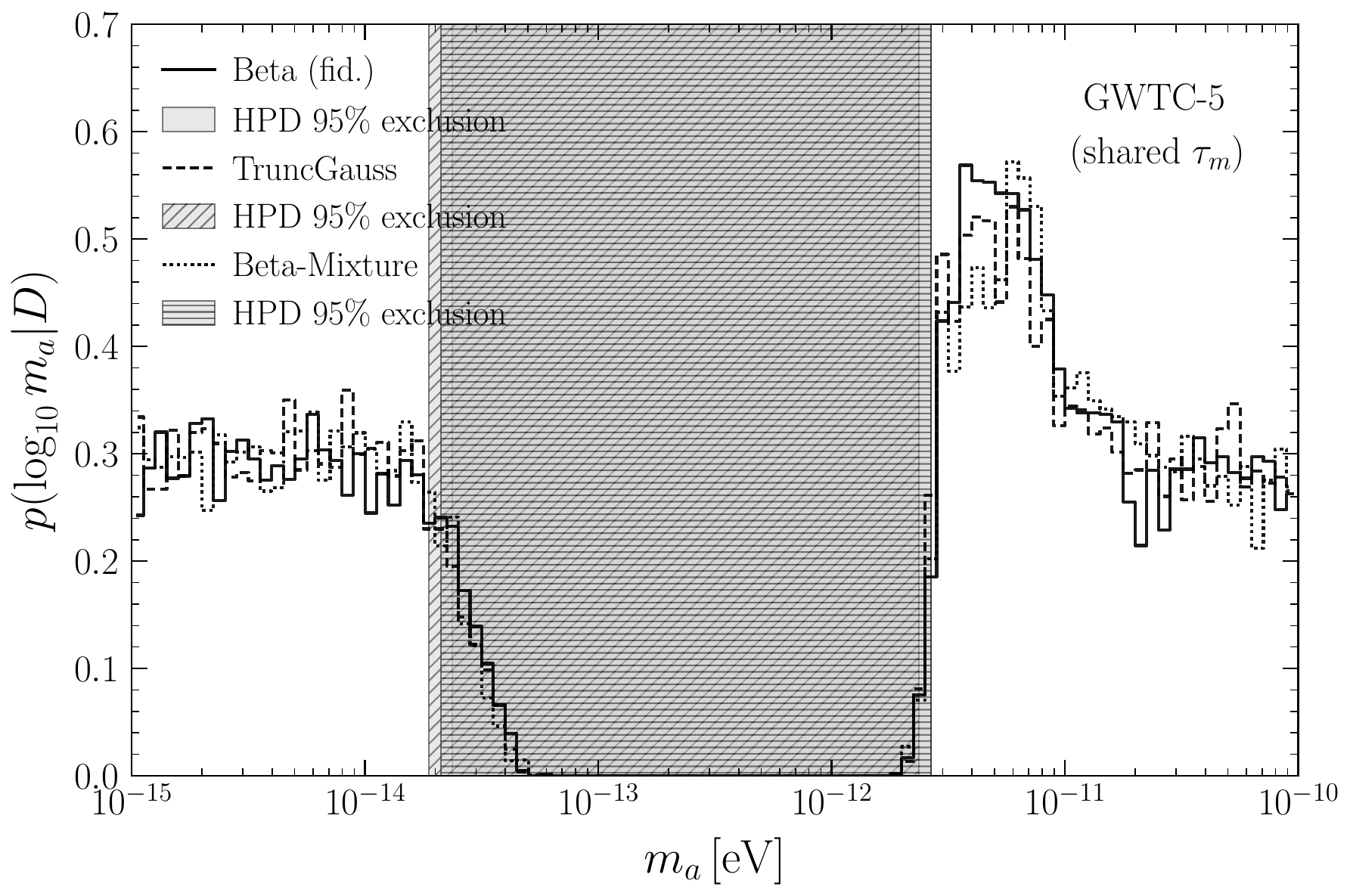}
\caption{(Left) Our fiducial axion mass $m_a$ marginalized posterior from GWTC-5 using the configuration where per-event $\tau_m$ are drawn from our DTD distribution, compared to the $m_a$ posteriors from an alternate simpler, but less realistic, analysis where all BBH events share a common singular $\tau_m$ from a log-flat prior $\tau_m \in [10^5, 10^9]$ yr. (Right) Our fiducial axion mass $m_a$ marginalized posterior from GWTC-5 using two other alternative natal spin distributions: a truncated gaussian and a Beta-mixture model (see text). Both result in similar $m_a$ posteriors as our fiducial Beta natal spin model. Note that we compare these alternative natal spin distributions assuming a single shared $\tau_m$ for computational simplicity, and to further isolate the effects of just changing the natal spin distribution.}
\label{fig:compare_tau_natal}
\end{figure}

In this section we analyze how our axion mass posterior results change under astrophysical systematic uncertainties in our modeling. First, in the left panel of Fig.~\ref{fig:compare_tau_natal} we compare our fiducial treatment of $\tau_m$ (where, for each BBH event, $\tau_m$ is drawn from a log-uniform DTD $p(\tau_m) \propto \tau_m^{-1}$ over $[10^7, \tau_H]$ yrs where $\tau_H$ is the Hubble time) to the simpler case of one shared $\tau_m$ amongst all BBH events assuming a log-uniform prior within the range $\tau_m \in [10^5,10^9]$ yr (motivated by the fact that Ref.~\cite{Ng:2020ruv} fixes $\tau_m = 10^7$ yr for all BBHs in their analysis). We find that the posteriors resulting from the shared $\tau_m$ scheme are slightly more conservative than our fiducial model, though they are less realistic, as one expects each BBH event to generically follow a unique $\tau_m$ given the multiple formation pathways available for BBH development~\cite{Wu:2024znq}.

Next, we test alternate natal spin distributions besides the fiducial single Beta distribution, which is a flexible two-parameter form that can capture the monotonic and single-peaked natal-spin distributions favored by current LVK inferences~\cite{KAGRA:2021duu,LIGOScientific:2025pvj,LIGOScientific:2026ctl}. We re-run the fiducial analysis under two additional models: a truncated Gaussian on $\chi \in [0, 1]$ with hyperparameters $(\mu_\chi, \sigma_\chi)$ for the mean and width (with uniform priors $\mu_{\chi} \in[0, 1]$ and $\sigma_{\chi} \in [0.05, 0.5])$, and a two-component Beta mixture with five hyperparameters $(\alpha_1, \beta_1, \alpha_2, \beta_2, w)$ specifying two Beta components and a mixing weight (the priors on $\alpha_i, \beta_i$ are the same as in the fiducial Beta model, and the mixing weight prior is taken as uniformly $w\in [0,1]$). The truncated Gaussian tests sensitivity to the specific Beta functional form at fixed unimodal location and width; the two-component mixture admits bimodal natal-spin distributions and distinct low- and high-spin sub-populations that a single Beta cannot represent. We apply these two additional models to our GWTC-5 analysis assuming a shared $\tau_m$ to isolate the effects of the natal spin model and for computational simplicity, finding that the $m_a$ posteriors and exclusion regions are robust under all three natal-spin models, see the right panel of Fig.~\ref{fig:compare_tau_natal}.

{\bf Mass-Dependent Distributions---}Recent population analyses find possible evidence that both the BBH DTD and the spin distribution evolve with mass, with a more rapidly spinning, plausibly hierarchical high-mass primary population emerging near the lower edge of the pair-instability gap at $\sim 45\,M_\odot$~\cite{Pierra:2024fbl, Tong:2025wpz}. To verify how our $m_a$ exclusion might be sensitive to this mass-dependence, we separately relax the mass-independence of the DTD and of the natal spin about the estimated split scale $M_s = 45\,M_\odot$. For the delay time, we decided the DTD per event by the median primary mass $\bar M_1$, holding low-mass events at the fiducial $p(\tau_m)\propto\tau_m^{-1}$ over $[10^7,\tau_H]$ and assigning high-mass events ($\bar M_1\ge M_s$) a modified power law $p(\tau_m)\propto\tau_m^{\alpha_\tau}$ over $[\tau_{\min}^{\rm hi},\tau_{\max}^{\rm hi}]$. We bracket the two physical directions motivated by formation-channel studies~\cite{vanSon:2021zpk, Padhyegurjar:2026slt}: (i) a shallower slope $\alpha_\tau=-0.35$, placing more weight at long delays so high-mass BHs have more time to spin down (pushing the low-$m_a$ edge lower); and, in contrast, (ii) a reduced floor $\tau_{\min}^{\rm hi}=10^5$~yr, allowing shorter delays and less time for SR.

\begin{figure}[htb]
\centering
\includegraphics[width=0.49\textwidth]{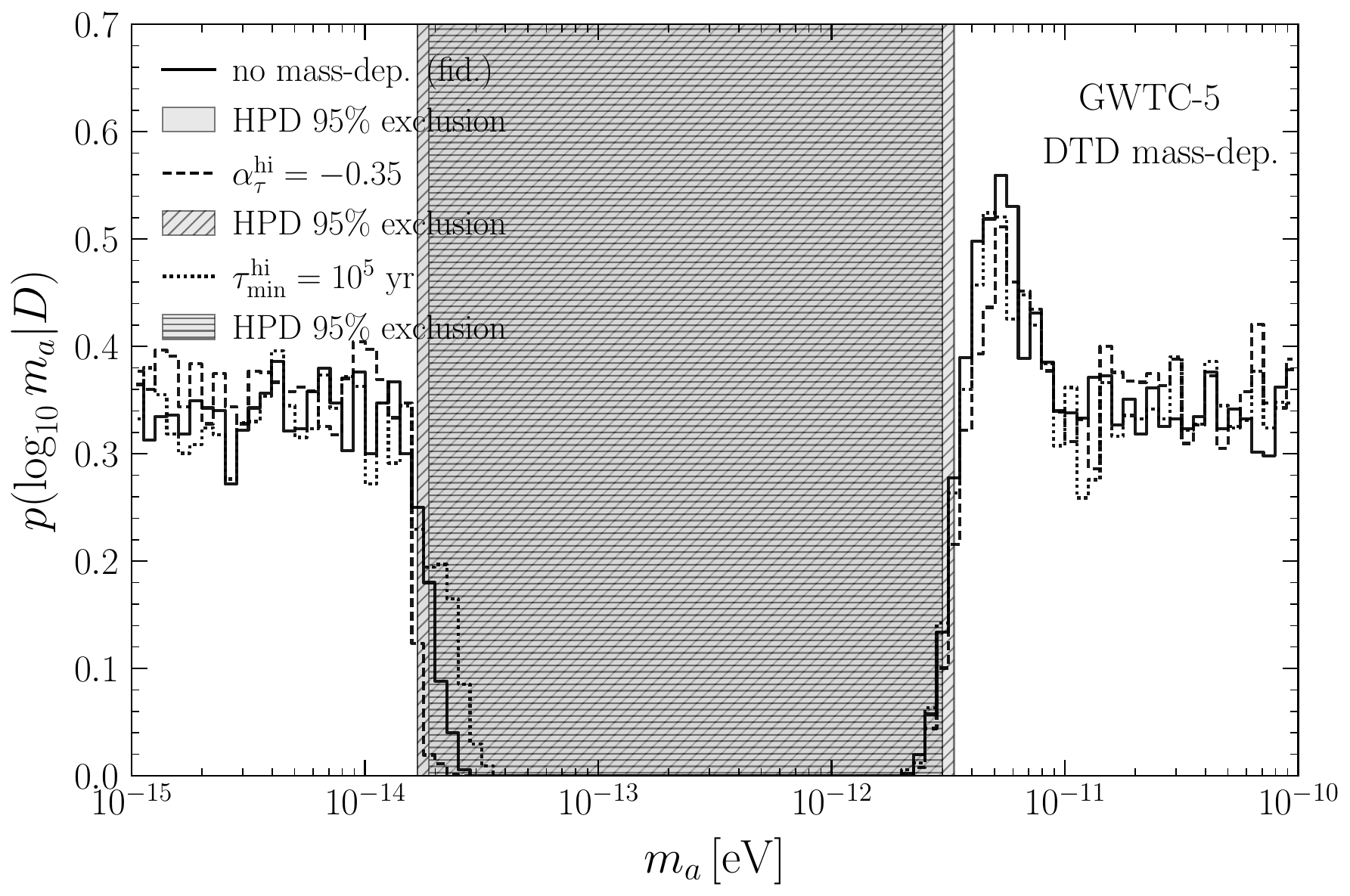}
\includegraphics[width=0.49\textwidth]{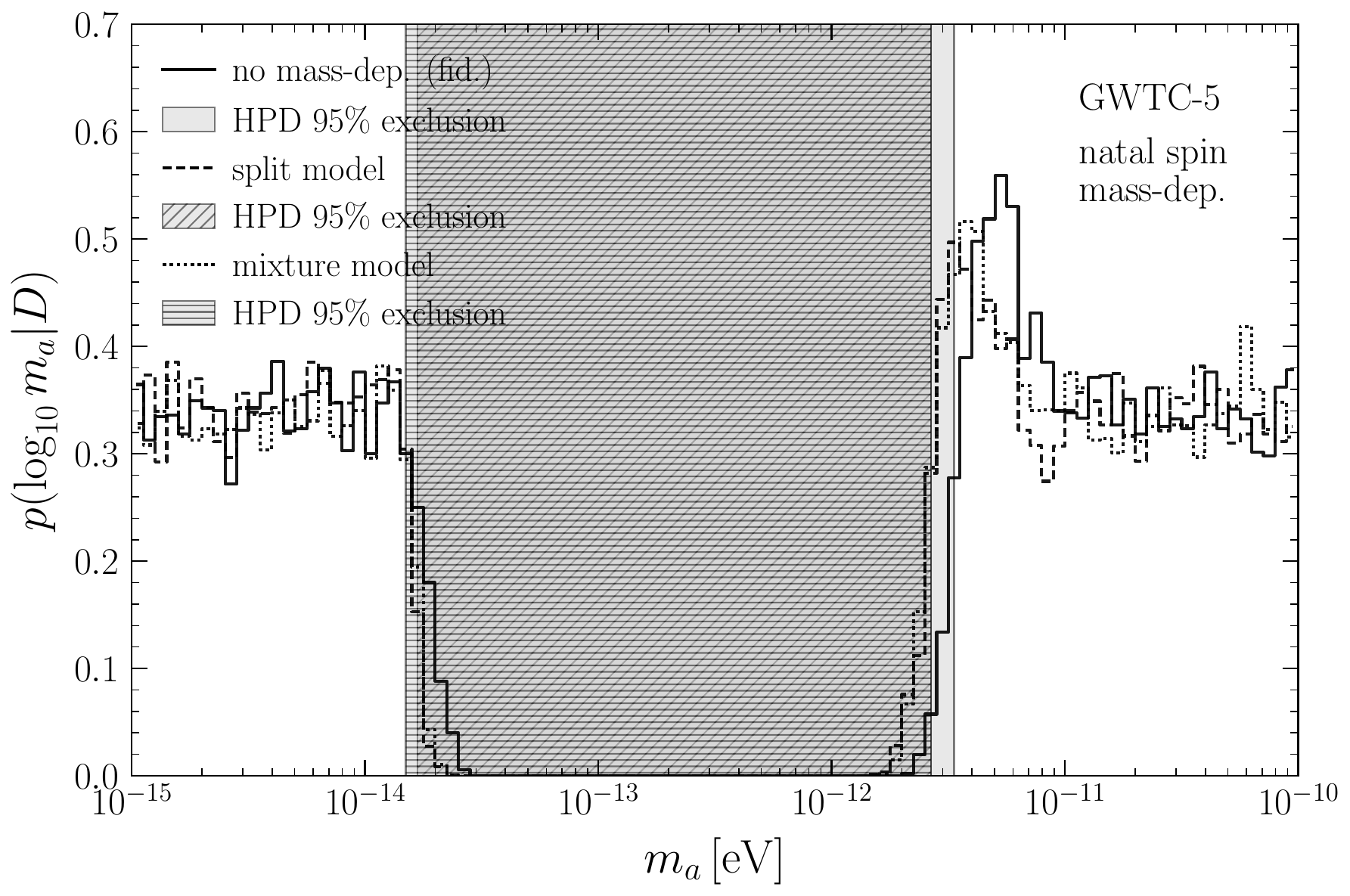}
\caption{(Left) The GWTC-5 axion mass posteriors across the two models we implement for a mass-dependent DTD (see main text), compared to our fiducial DTD model (with no mass-dependence). (Right) The same but comparing the posteriors obtained from our two models of mass-dependent natal spin distributions.}
\label{fig:compare_mass_dep}
\end{figure}

For the natal spin, we modify our Beta models to take into account, per PE sample, the primary mass-dependence, in two forms. The first (which we term the `split' model) involves two independent Betas across the split, $(\alpha,\beta)(M)=(\alpha_{\rm lo},\beta_{\rm lo})$ for $M<M_s$ and $(\alpha_{\rm hi},\beta_{\rm hi})$ for $M\ge M_s$ (fiducial Beta priors on each bin). Note that this is the same as the two-component Beta mixture model discussed more generically earlier, but here explicitly dependent on mass. The second (which we term the `mixture' model) involves a mass-conditioned two-component mixture of the form~\cite{Berti:2025usa}
\begin{equation}
p(\chi\mid M)=\big[1-w(M)\big]\,{\rm Beta}(\chi;\alpha_1,\beta_1)
+w(M)\,{\rm Beta}(\chi;\alpha_2,\beta_2),\qquad
w(M)=\frac{w_{\max}}{1+e^{-(M-M_s)/\Delta}},
\end{equation}
where we note that the limit $w_{\max}\to0$ recovers the fiducial single Beta distribution. Across all DTD and natal-spin variants and all $M_s$, the $m_a$ marginal posterior and its $95\%$ HPD exclusion are roughly comparable with the fiducial result, noting that the $m_a$ exclusions from both mass-dependent spin distributions have upper edges that are slightly lower than our fiducial model. We show these comparisons in the left and right panels of Fig.~\ref{fig:compare_mass_dep} for the mass-dependent delay time and natal spin analyses, respectively. We note that while more elaborate mass-dependent parameterizations of the DTD and natal spin distributions are possible, all parameterizations carry their own astrophysical uncertainties that are themselves still being constrained by current and future gravitational wave data. The relative consistency of our exclusion across the variants tested above suggests that our mass-independent fiducial remains a reasonable choice given present uncertainties.

{\bf Tidal Perturbations---}A superradiant cloud around one binary component can be tidally perturbed by its companion, which mixes the cloud's bound state with other superradiant levels and, through absorptive partner states, can deplete the cloud before merger~\cite{Zhu:2024bqs, Caputo:2025oap}. We test for these possible effects following the level-mixing formalism of Ref.~\cite{Caputo:2025oap}, where we compute the tidally-induced rate shift $\delta\Gamma_{n\ell m}$ and define survival by $\Gamma_{n\ell m}+\delta\Gamma_{n\ell m}>0$; the selection rules make the $m=1$ states robust, while $m=2$ can couple to, {\it e.g.}, absorptive channels and is vulnerable. Following Ref.~\cite{Caputo:2025oap} we sum all allowed partners up to $m=2$ (with higher modes increasingly inhibitive to properly compute), noting that the survival condition is equivalent to a coupling threshold $\alpha>\alpha_{\rm crit}(M,M_*,\chi,\tau_{\rm cmp})$ where $M_*$ is the mass of the companion; if a BH cannot satisfy this threshold we do not include it for this systematic analysis.

\begin{figure}[htb]
\centering
\includegraphics[width=0.5\textwidth]{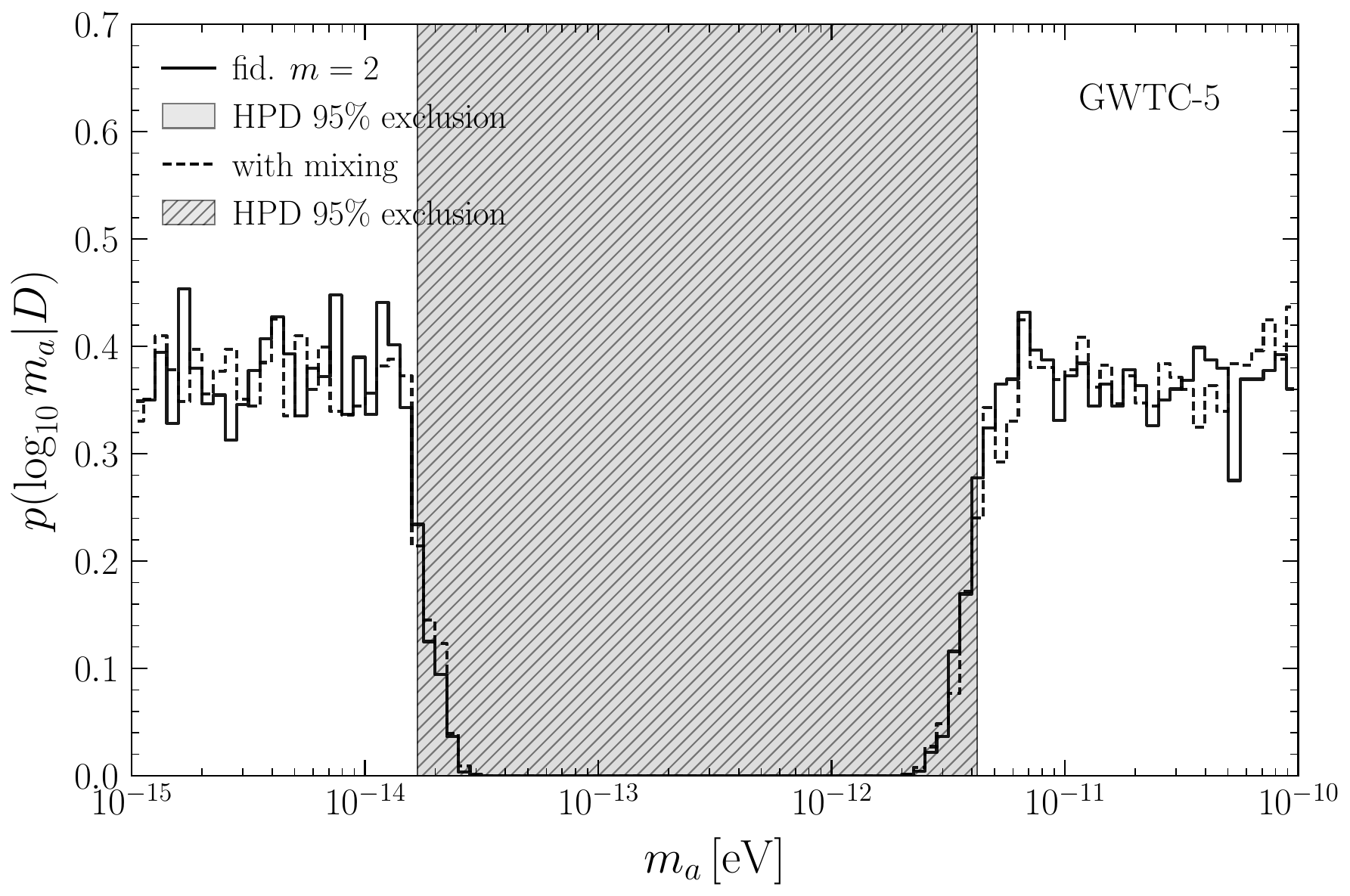}
\caption{A comparison of the axion mass posteriors from a $m=2$ GWTC-5 analysis with and without the effects of mixing from tidal perturbations. We note the minimal effects of mixing in the final posteriors.}
\label{fig:compare_tidal}
\end{figure}

We incorporate these mixing effects into our GWTC-5 search, and compare to our GWTC-5 analysis, limited to $m=2$ for comparison, in Fig.~\ref{fig:compare_tidal}. We find that the posteriors on $m_a$ are largely unchanged; the tidal disruption is expected to potentially affect the small $\alpha$ regime of low $m_a$, though our low-$m_a$ edge is set by the most massive events, whose $\alpha$ there already exceeds $\alpha_{\rm crit}$. Hence, we find that the effects of tidal disruption, at least qualified up to the $m=2$ implementation here, are likely minimal for our analyses, though a full characterization up to arbitrary modes would be an avenue for future work.

\begin{figure}[htb]
\centering
\includegraphics[width=0.5\textwidth]{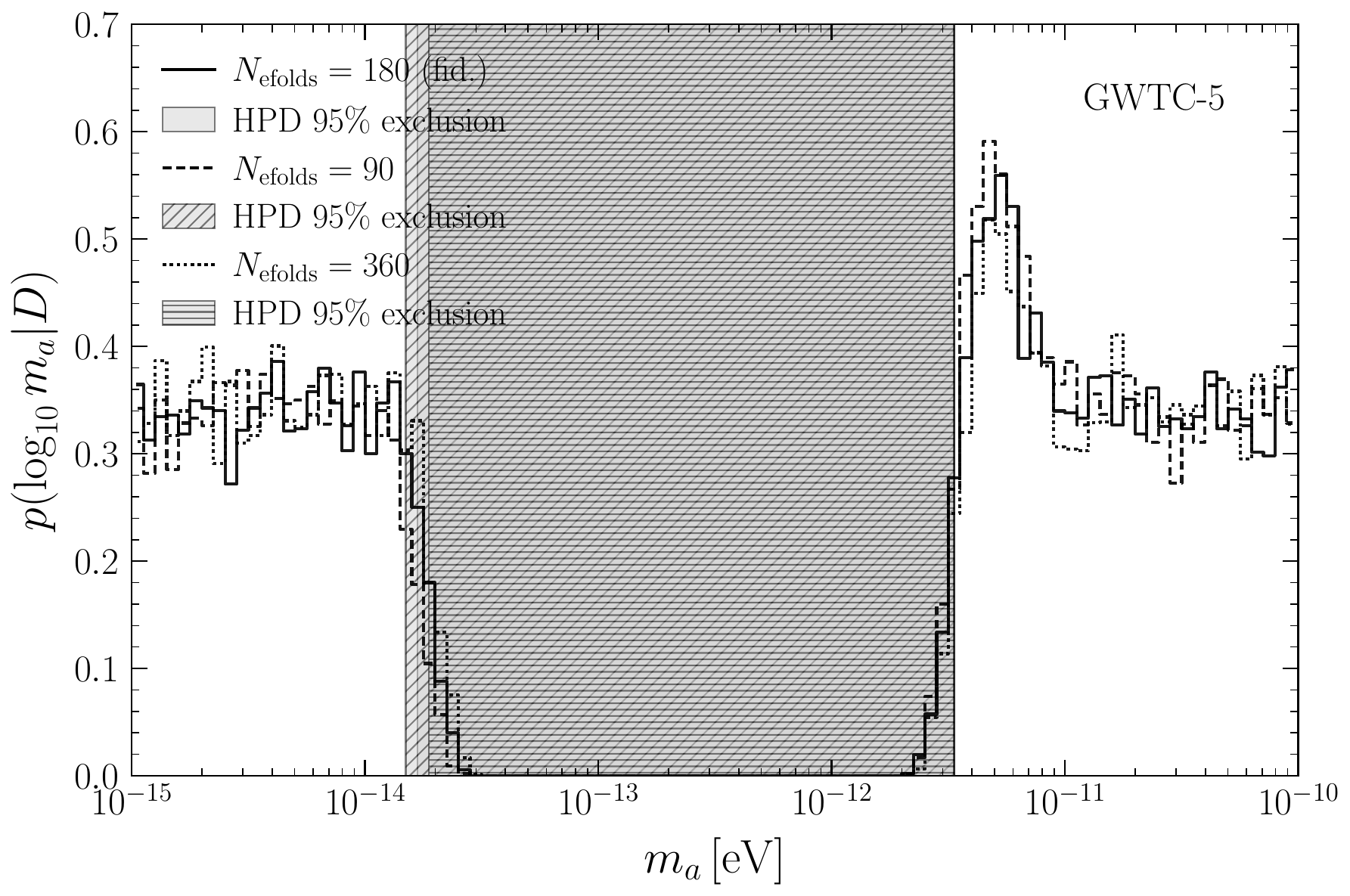}
\caption{Our fiducial axion mass $m_a$ marginalized posterior from GWTC-5 under different $N_{\rm efolds}$ requirements for SR cloud growth. All result in similar $m_a$ posteriors compared to our fiducial $N_{\rm efolds} = 180$ benchmark.}
\label{fig:compare_Nefolds}
\end{figure}

{\bf Number of SR e-folds---}Finally, we test how our fiducial results change when a different number of $N_{\rm efolds}$ is used when assuming axion cloud growth around the BH via SR. In our fiducial model we assume $N_{\rm efolds} = 180$ as estimated in Refs.~\cite{Arvanitaki:2010sy, Arvanitaki:2014wva, Ng:2019jsx}, and we bracket our uncertainties to half and double this quantity (which is then exponentiated in the context of the saturation occupation number itself). Running our fiducial analysis for GWTC-5 for these two uncertainties result in little change to our ultimate $m_a$ posteriors and exclusion regions, see Fig.~\ref{fig:compare_Nefolds}. The minor discrepancy is expected, where we see that higher (lower) $N_{\rm efolds}$ makes SR growth more (less) restrictive, which is marginally reflected in our resulting posteriors.

\end{document}